\documentclass[twocolumn]{aastex63}
\usepackage{graphicx} 
\usepackage{threeparttable} 
\usepackage[caption=false]{subfig}
\usepackage{graphicx}
\usepackage{multirow}
\usepackage{graphicx}
\usepackage{booktabs}
\usepackage{amsmath,amssymb}
\usepackage[T1]{fontenc}
\usepackage{epstopdf}
\graphicspath{{./}{figures/}}

\begin{document}

\title{Fermi Large Area Telescope Detection of Gamma-Rays from NGC 6251 Radio Lobe}

\correspondingauthor{Jin Zhang}
\email{j.zhang@bit.edu.cn}

\author[0009-0000-6577-1488]{Yu-Wei Yu}
\affiliation{School of Physics, Beijing Institute of Technology, Beijing 100081, People's Republic of China; j.zhang@bit.edu.cn}

\author[0000-0001-6863-5369]{Hai-Ming Zhang}
\affiliation{School of Astronomy and Space Science, Nanjing University, Nanjing 210023, People's Republic of China}

\author[0000-0002-4789-7703]{Ying-Ying Gan}
\affiliation{School of Physics, Beijing Institute of Technology, Beijing 100081, People's Republic of China; j.zhang@bit.edu.cn}

\author{Xin-Ke Hu}
\affiliation{Guangxi Key Laboratory for Relativistic Astrophysics, School of Physical Science and Technology, Guangxi University, Nanning 530004, People's Republic of China}

\author{Tan-Zheng Wu}
\affiliation{School of Physics, Beijing Institute of Technology, Beijing 100081, People's Republic of China; j.zhang@bit.edu.cn}

\author[0000-0003-3554-2996]{Jin Zhang\dag}
\affiliation{School of Physics, Beijing Institute of Technology, Beijing 100081, People's Republic of China; j.zhang@bit.edu.cn}

\begin{abstract}

We report on the detection of extended $\gamma$-ray emission from lobes in the radio galaxy NGC 6251 using observation data from Fermi Large Area Telescope (Fermi-LAT). The maximum likelihood analysis results show that a radio morphology template provides a better fit than a point-like source description for the observational data at a confidence level of 8.1$\sigma$, and the contribution of lobes constitutes more than 50\% of the total $\gamma$-ray flux. Furthermore, the $\gamma$-ray energy spectra show a significant disparity in shape between the core and lobe regions, with a curved log-parabola shape observed in core region and a power-law form observed in lobes. Neither the core region nor the northwest lobe displays significant flux variations in the long-term $\gamma$-ray light curves. The broadband spectral energy distributions of both the core region and northwest lobe can be explained with a single-zone leptonic model. The $\gamma$-rays of the core region are due to the synchrotron-self-Compton process while the $\gamma$-rays from northwest lobe are interpreted as inverse Compton emission of the cosmic microwave background. 

\end{abstract}


\section{Introduction}

Radio galaxies (RGs) are a subclass of radio-loud (RL) active galactic nuclei (AGNs), characterized by the remarkable and complex large-scale jet morphology in the radio band. The large-scale jet can extend beyond the host galaxy of RGs and form some substructures, such as knots, hotspots, and radio lobes, which have been resolved in the radio, optical, and X-ray bands (see \citealp{2006ARA&A..44..463H} for a review). So far, dozens of RGs have been detected by the Fermi Large Area Telescope (Fermi-LAT, \citealp{2023arXiv230712546B,2022ApJS..260...53A}). Several RGs are even detected at the TeV band and confirmed to be very high energy emitters\footnote{http://tevcat.uchicago.edu/}. Generally, the $\gamma$-rays of RGs are still considered to come from the core-jet \citep{2015ApJ...798...74F,2017RAA....17...90X}. However, some works studying the X-ray emission of the large-scale jet substructures in RGs predicted the detectable $\gamma$-rays of some substructures (e.g., \citealp{2009ApJ...701..423Z,2010ApJ...710.1017Z,2018ApJ...858...27Z}). Especially, the detection of $\gamma$-rays from large-scale radio lobes of RGs Cen A \citep{2010Sci...328..725A} and Fornax A \citep{2016ApJ...826....1A} confirms that the substructures of jet are acceleration sites of high energy particles. Hence, the origin of $\gamma$-ray emission for RGs is still debated. 

NGC 6251 is a nearby (z=0.0247, \citealp{2003AJ....126.2268W}) giant RG with a radio angular scale extending to about $1^{\circ}.2$. Its radio morphology mainly consists of a radio core, a narrow straight jet with roughly uniform cone angle, and two large-scale radio lobes \citep{1977MNRAS.181..465W,1981MNRAS.197..287S}. The jet starts within 1 pc of the nucleus, extends towards the northwest (NW) for approximately $4^{\prime}.5$, and subsequently forms the NW lobe about $0^{\circ}.3$ away from the radio core \citep{1981MNRAS.197..287S,1984ApJS...54..291P}. The radio intensities of both the jet and NW lobe are clearly higher than that of the southeast (SE) lobe, which is separated from the radio core by an angular distance of $0^{\circ}.7$ \citep{1984ApJS...54..291P,2020MNRAS.495..143C}. The X-rays from the nucleus and its $\sim4^{\prime}.5$ jet were initially observed by ROSAT \citep{1993ApJ...412..568B}. The diffuse X-ray emission originating from the NW lobe in NGC 6251 has been detected through the Suzaku X-ray imaging observations, which was suggested to be generated by the inverse Compton scattering of cosmic microwave background (IC/CMB) process \citep{2012ApJ...749...66T}. The X-ray emission from the straight jet region was also reported to potentially arise from the IC/CMB process \citep{2004A&A...414..885S}, although an alternative explanation could be synchrotron radiation \citep{2005MNRAS.359..363E}.

Gamma-rays from NGC 6251 were likely first detected by the Compton Gamma-Ray Observatory in the MeV-GeV range. Using the X-ray images obtained from ROSAT and ASCA observations, \cite{2002ApJ...574..693M} proposed that the Energetic Gamma-Ray Experiment Telescope source 3EG J1621+8203 is likely associated with NGC 6251. This association was further substantiated by \cite{2005A&A...433..515F} through an analysis a wide-view-field observation from INTEGRAL/IBIS. The GeV $\gamma$-ray emission of NGC 6251 has been reported in the first Fermi-LAT catalog \citep{2010ApJS..188..405A}, identified as a $\gamma$-ray emitting misaligned AGN \citep{2010ApJ...720..912A}, and subsequently included in the second/third/fourth Fermi-LAT catalogs \citep{2012ApJS..199...31N,2015ApJS..218...23A,2022ApJS..260...53A,2023arXiv230712546B}. The observed $\gamma$-rays are conventionally believed to originate from the core-jet region near the black hole via the synchrotron-self-Compton (SSC) process \citep{2003ApJ...597..166C,2011A&A...533A..72M,2017RAA....17...90X}. However, the 95\% error ellipse of its associated Fermi-LAT source 2FGL J1629.4+8236 does not encompass the radio core of NGC 6251 \citep{2012ApJ...749...66T}. Additionally, the corresponding coordinates of this Fermi-LAT source in the latest fourth Fermi-LAT source catalog fall within the region of the jet in NGC 6251, being at a distance of $\sim3^{\prime}.6$ from the radio core \citep{2022Univ....8..587F}. These results make the origin of its $\gamma$-rays ambiguous. In this paper, we completely analyze the $\sim$15-yr Fermi-LAT observation data to investigate the origins of its $\gamma$-ray emission. The data analysis and results of Fermi-LAT are presented in Section \ref{sec2}. The spectral energy distributions (SEDs) of NW lobe and core region are constructed and modelled in Section \ref{sec3}. Discussion and conclusions are given in Section \ref{sec4}. Throughout, $H_0=71$ km s$^{-1}$ Mpc$^{-1}$, $\Omega_{\rm m}=0.27$, and $\Omega_{\Lambda}=0.73$ are adopted in this paper.

\section{Fermi-LAT Observations and Data Analysis}
\label{sec2}
\subsection{Baseline Model of Data Analysis} 

NGC 6251 has been reported to be associated with a $\gamma$-ray source 4FGL 1630.6+8234 in the Fermi-LAT 12-yr Source Catalog (4FGL-DR3, \citealp{2022ApJS..260...53A})\footnote{We have checked the Fermi-LAT latest Source Catalog (4FGL-DR4, \citealp{2023arXiv230712546B}), the results are consistent.}. The Pass 8 data covering from 2008 August 4 to 2023 March 1 (MJD 54682--60004) within the energy range of 0.1--300 GeV were downloaded from the Fermi Science Support Center\footnote{https://fermi.gsfc.nasa.gov/cgi-bin/ssc/LAT/LATDataQuery.cgi}. We select the data within 15$^\circ$ region of interest (ROI) centered on the radio position of NGC 6251 (R.A.=248$^\circ$.133, decl.=82$^\circ$.538). The publicly available software \textit{fermitools} (ver. 2.2.0, \citealp{2019ascl.soft05011F}) and the binned likelihood analysis method are used for our analysis. We use event class ``SOURCE'' (evclass=128) and event type ``FRONT+BACK'' (evtype=3) for data analysis based on LAT data selection recommendations. To eliminate the contamination of $\gamma$-rays from the Earth limb, the maximum zenith angle of 90$^\circ$ is set. A standard filter expression ``(DATA\_QUAL\textgreater0)\&\&(LAT\_CONFIG==1)'' and the instrument response function of $\rm P8R3\_SOURCE\_V3$ are used. All sources included in the 4FGL-DR3 within ROI are added to the model. The spectral parameters of the sources encompassed in the circle centered on 4FGL J1630.6+8234 with a radius of $6^\circ$ are left free, whereas the parameters of those sources lying beyond $6^\circ$ are fixed to their 4FGL-DR3 values. The background models including isotropic emission (``$\rm iso\_P8R3\_SOURCE\_V3\_V1.txt$'') and the diffuse Galactic interstellar emission (``$\rm gll\_iem\_v07.fits$'') are considered and only their normalization parameters are kept free.

We use the maximum test statistic (TS) to evaluate the significance of the $\gamma$-ray signals of a source in the background, $\rm TS=2(log\mathcal{L}_{\rm src}-log\mathcal{L}_{\rm null})$, where $\mathcal{L}_{\rm src}$ and $\mathcal{L}_{\rm null}$ are the likelihood values for the background with and without a source. If TS$\geq$25, it is considered that there is a new source \citep{2010ApJS..188..405A}. By generating a $5^\circ\times5^\circ$ residual TS map centered on the radio core of NGC 6251 for new source search, the maximum TS value of $\sim$10 is obtained, indicating that no new sources appear in the background. Generally, the power-law (PL) or log-parabola (LP) functions are used to fit the spectrum of sources, i.e., $dN(E)/dE=N_0(E/E_{\rm b})^{-\Gamma_{\gamma}}$ or $dN(E)/dE=N_0(E/E_{\rm b})^{-(\Gamma_{\gamma}+\beta\log(E/E_{\rm b}))}$, where $N(E)$ is the photon distribution as a function of energy, $\Gamma_{\gamma}$ is the photon spectral index, $E_{\rm b}$ is the scale parameter of photon energy, and $\beta$ is the curvature parameter \citep{2004A&A...413..489M}. Following the methodology outlined in the second Fermi-LAT catalog \citep{2012ApJS..199...31N}, we first calculate $\rm TS_{\rm curve}=2(log\mathcal{L}_{\rm LP}-log\mathcal{L}_{\rm PL})$ to estimate the curvature significance of an energy spectrum. If $\rm TS_{\rm curve}\geq16$ (corresponding to 4$\sigma$), it indicates that the spectrum exhibits significant curvature, and therefore we use the LP function to describe it.

\subsection{Point-Like Source Analysis} 

\begin{figure}
 \centering
 \includegraphics[angle=0,scale=0.24]{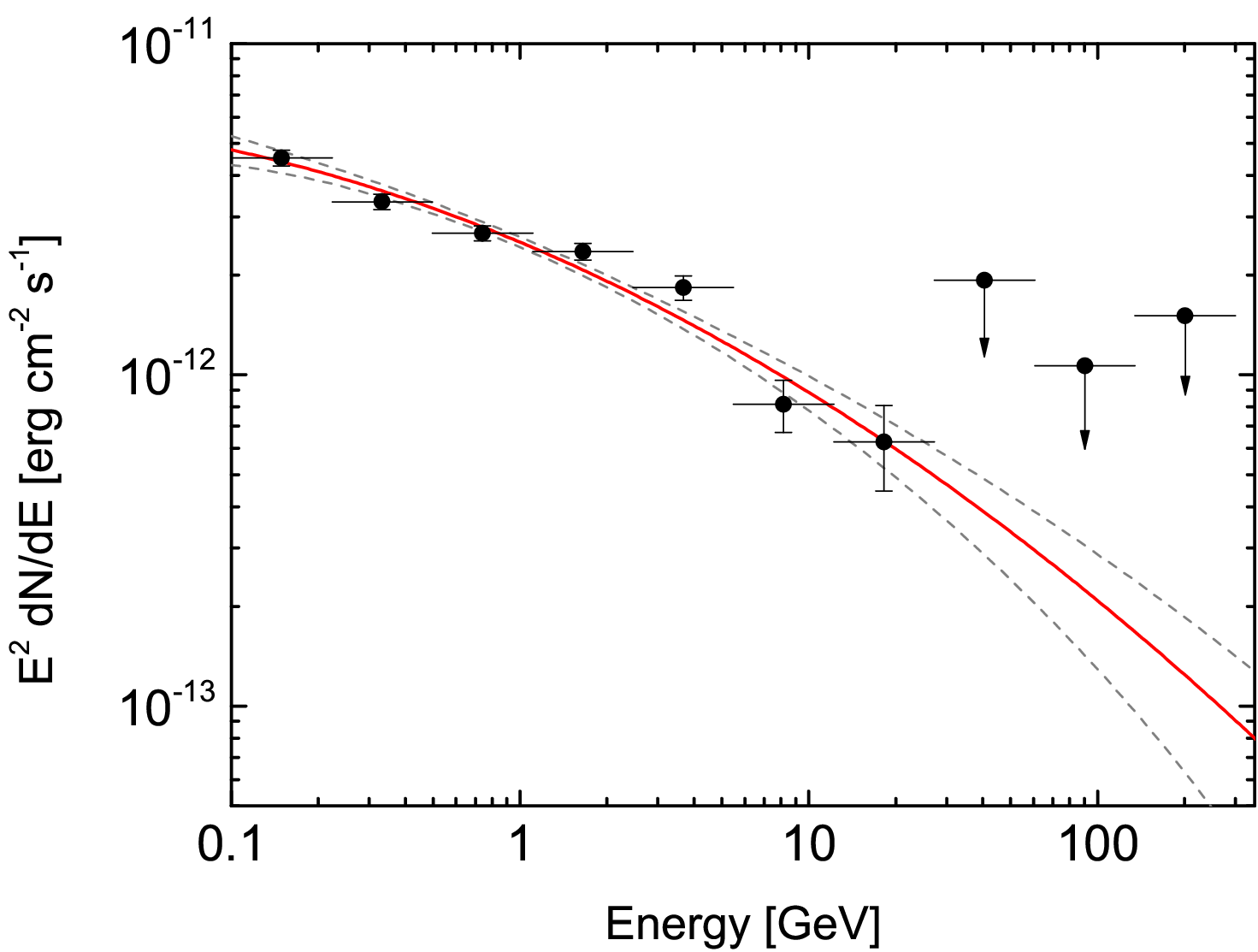}
 \includegraphics[angle=0,scale=0.38]{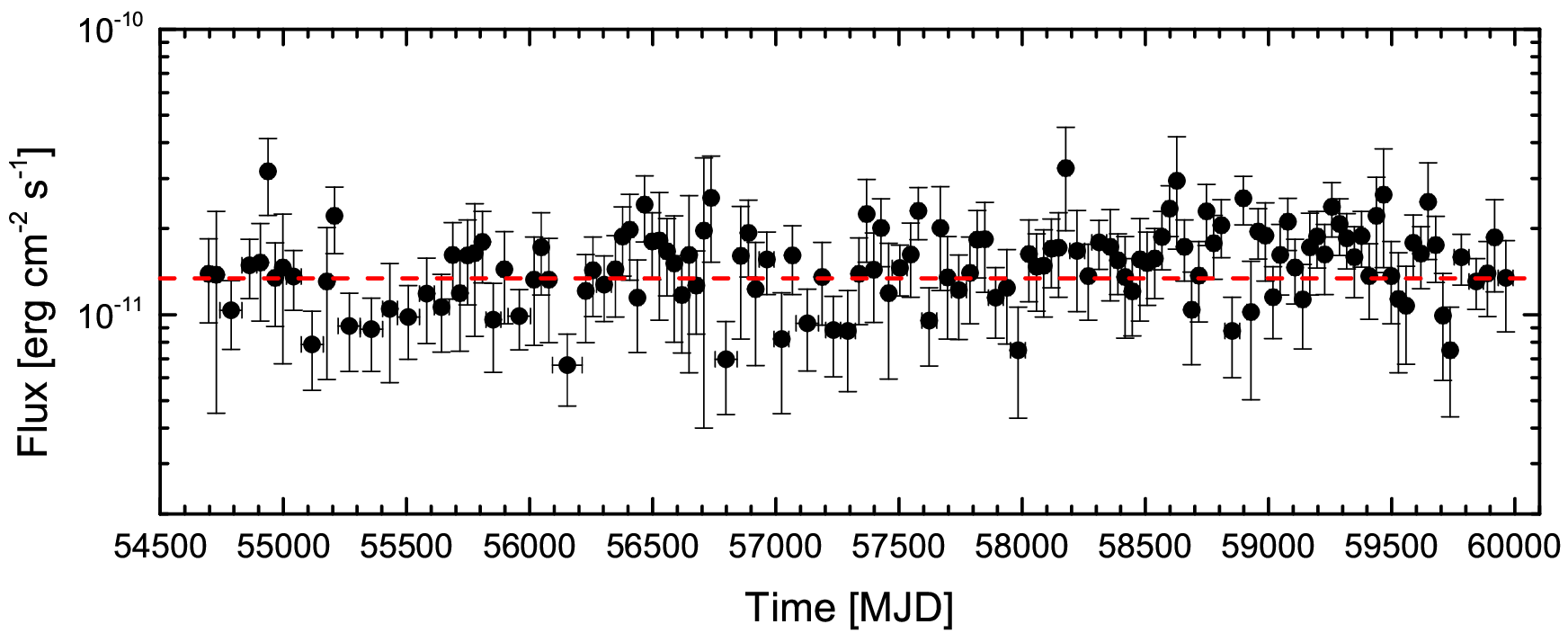}
\caption{Spectrum and light curve of 4FGL J1630.6+8234 when considering it as a PS. They are all derived with the $\sim$15-yr Fermi-LAT observation data in the 0.1--300 GeV band. The solid red line and gray dash lines (in left panel) represent the fitting result of the spectrum and the corresponding 1$\sigma$ uncertainty, respectively. If TS$<$4, an upper-limit is presented for that energy bin. The light curve (in right panel) is obtained with an adaptive-binning method based on a criterion of TS$\ge$9 for each time bin, where the minimum time-bin step is 30 days. The horizontal red dashed line is the $\sim$15-yr average flux, i.e., $F_{\rm 0.1-300~GeV}=(13.40\pm0.51)\times10^{-12}$~erg~cm$^{-2}$~s$^{-1}$.}\label{point}
\end{figure}

In the 4FGL-DR4 \citep{2023arXiv230712546B,2022ApJS..260...53A}, 4FGL J1630.6+8234 is identified as a point-like source (PS) with a LP spectrum. Reanalyzing its $\sim$15-yr Fermi-LAT observation data, the TS map also shows it as a PS with TS$\sim$2619.34. Its time-integrated spectrum of the $\sim$15-yr Fermi-LAT observations within the energy band of 0.1--300 GeV is well explained by a LP spectral form with $\Gamma_{\gamma}=2.36\pm0.03$ and $\beta=0.04\pm0.02$, as depicted in Figure \ref{point} and shown in Table \ref{tab:model}. The $\sim$15-yr average flux is $F_{\rm 0.1-300~GeV}=(13.40\pm0.51)\times10^{-12}$ erg cm$^{-2}$ s$^{-1}$. These results are roughly consistent with that of $\Gamma_{\gamma}=2.27\pm0.04$, $\beta=0.10\pm0.02$, and $F_{\rm 0.1-100~GeV}=(13.10\pm0.56)\times10^{-12}$ erg cm$^{-2}$ s$^{-1}$ in the 4FGL-DR4. The long-term light curve of 4FGL J1630.6+8234 is generated using the adaptive-binning method \citep{2012A&A...544A...6L} to ensure TS$\ge$9 for each time bin, where the minimum bin size is 30 days. The $\sim$15-yr Fermi-LAT light curve of 4FGL J1630.6+8234 in the 0.1--300 GeV band demonstrates a steady state without any significant flux variation, as displayed in Figure \ref{point}. 

Using the tool of $gtfindsrc$, we re-estimate the best-fit position of 4FGL J1630.6+8234 and obtain (R.A.=247$^\circ$.752, decl.=82$^\circ$.558) with a 95\% confidence error circle radius of 0$^\circ$.015. This position is located in the northwest region relative to the radio core of NGC 6251, as presented in Figure \ref{countsmap}. Together with the steady $\gamma$-ray emission of 4FGL J1630.6+8234 in the past $\sim$15-yr, it indicates that the $\gamma$-ray emission of 4FGL J1630.6+8234, at least a portion of $\gamma$-rays, is likely dominated by the radiation of the large-scale extended regions in NGC 6251.

\subsection{Radio Morphology Template}

In order to examine our speculation, we firstly generate the $\gamma$-ray counts map of 4FGL J1630.6+8234 in the 0.1--300 GeV band, as depicted in Figure \ref{countsmap}. The data reduction procedure is described in the Appendix. The radio contours of NGC 6251 at 609 MHz observed by the Westerbork Synthesis Radio Telescope (WSRT) \citep{1997A&AS..123..423M}, the 95\% error ellipses for the counterpart of NGC 6251 from 1FGL to 4FGL, and the 95\% error circle of the best-fit position of this work are also presented in Figure \ref{countsmap}. The confidence intervals of the best-fit position decrease with increasing LAT observation time, but the radio core of NGC 6251 is never encompassed. The distribution of $\gamma$-ray photons on the NW lobe of NGC 6251 is significantly prominent, while the brightness of the SE lobe appears relatively subdued, rendering it indistinguishable in the map. As shown in Figure \ref{countsmap4} in the Appendix, the spatial distributions of $\gamma$-ray photons exhibit significant variations across the low-energy to high-energy bands, shifting from a diffusion distribution dominated by the large-scale extended regions to a PS dominated by the radio core. The $\gamma$-ray photons above 10 GeV are primarily concentrated in close proximity to the radio core region. 

Assuming that $\gamma$-rays are produced through the IC process by the same group of electrons responsible for generating radio radiation, the distribution of $\gamma$-rays can be considered as a form of radio radiation \citep{2016ApJ...826....1A}. Therefore, using a Radio Morphology Template (RMT) model, we reanalyze the $\sim$15-yr Fermi-LAT observation data of 4FGL J1630.6+8234. We employ RMT to partition the observation area into three zones, namely, the core region and two giant lobes. The regions of two giant lobes are selected with the 609 MHz radio image from the \citet{NED}. Of particular note, the overlap region between the radio green contours and the red circle (Figure \ref{countsmap4} in the Appendix) is subtracted and not included for the NW lobe template in the RMT model. The red circle is centered on the radio core with a radius of $4^{\prime}.5$, which encompasses the radio core, a bright straight jet extending to the NW from the radio core, and the faint counterjet. The two lobes are taken as extended radiation sources while the core region is set as a PS model centered around the radio core of NGC 6251. We use the RMT model instead of the PS 4FGL J1630.6+8234 in the XML file and re-perform the maximum likelihood analysis. 

\begin{figure}
 \centering
 \includegraphics[angle=0,scale=0.24]{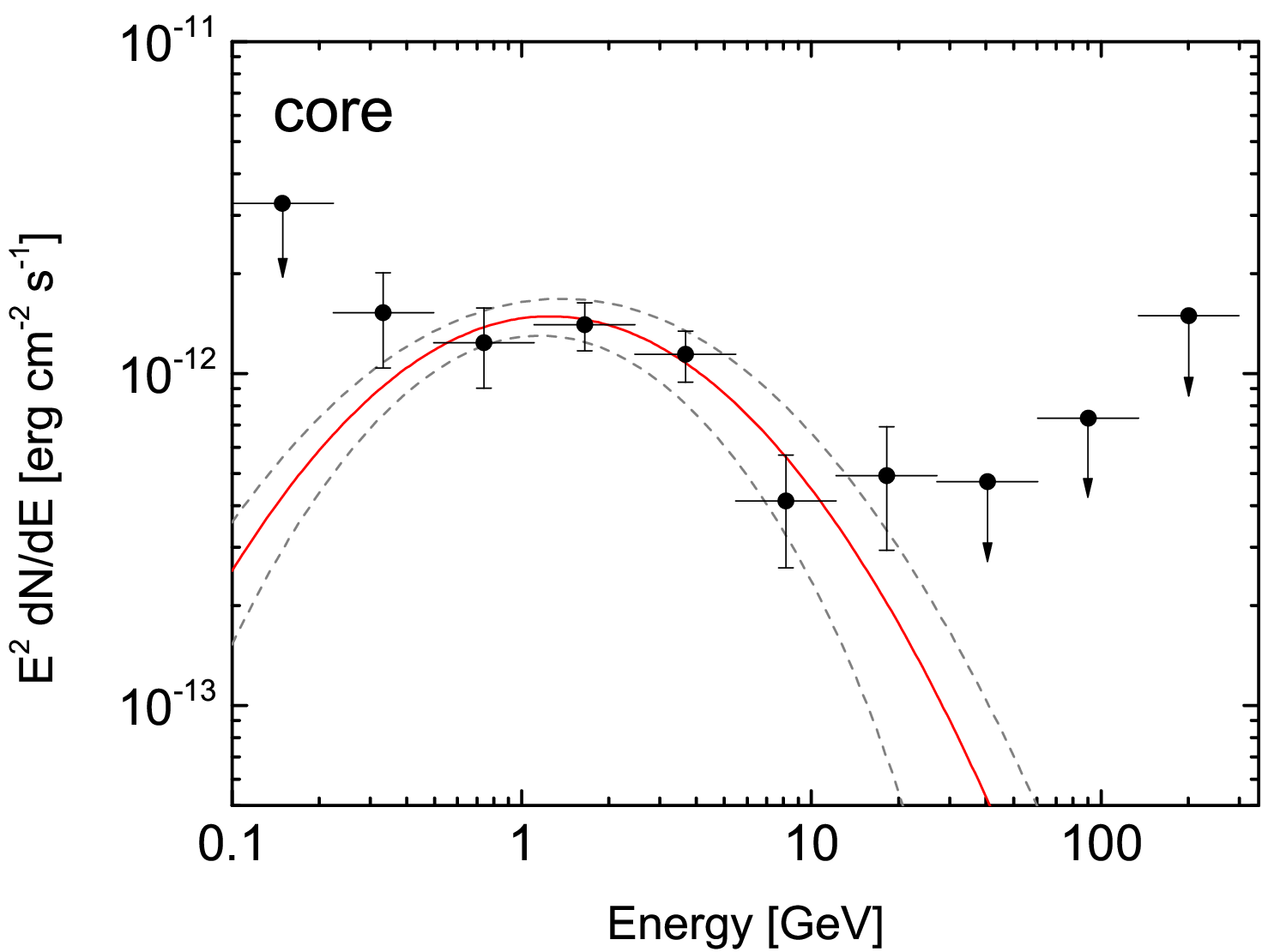}
 \includegraphics[angle=0,scale=0.38]{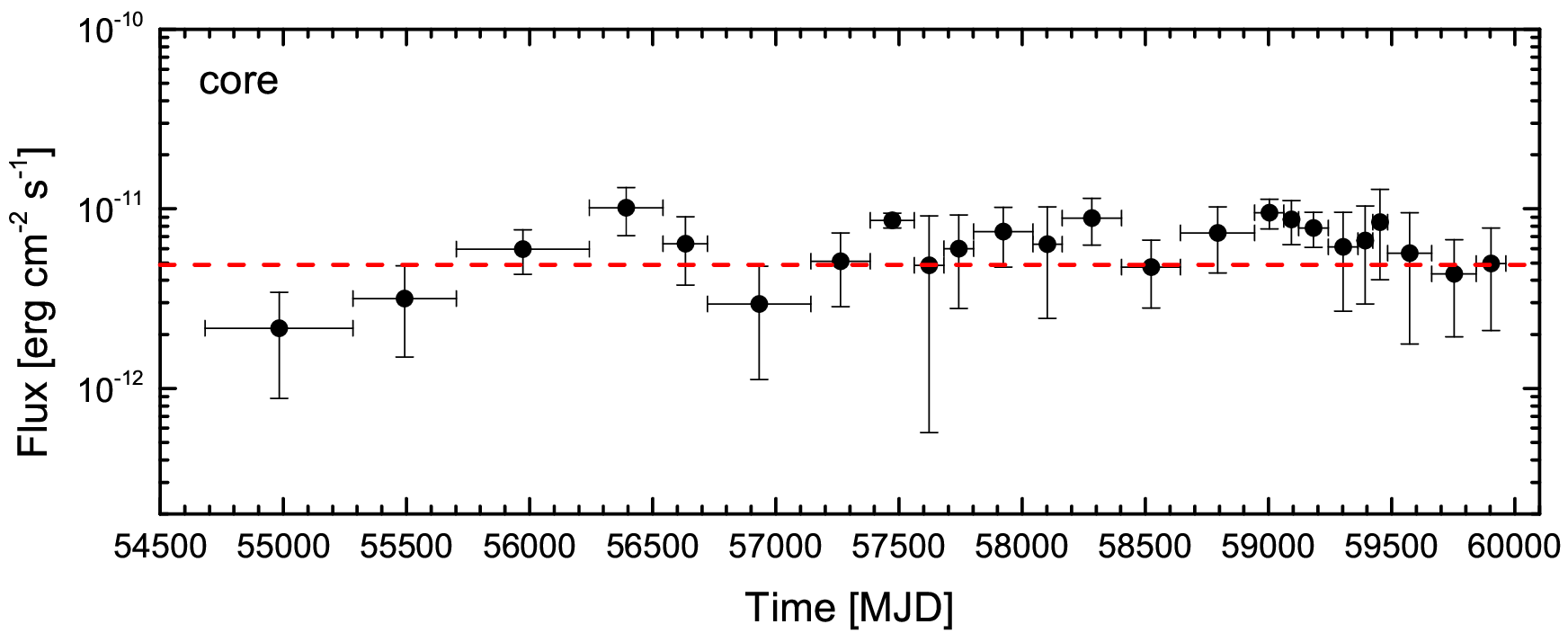}
 \includegraphics[angle=0,scale=0.24]{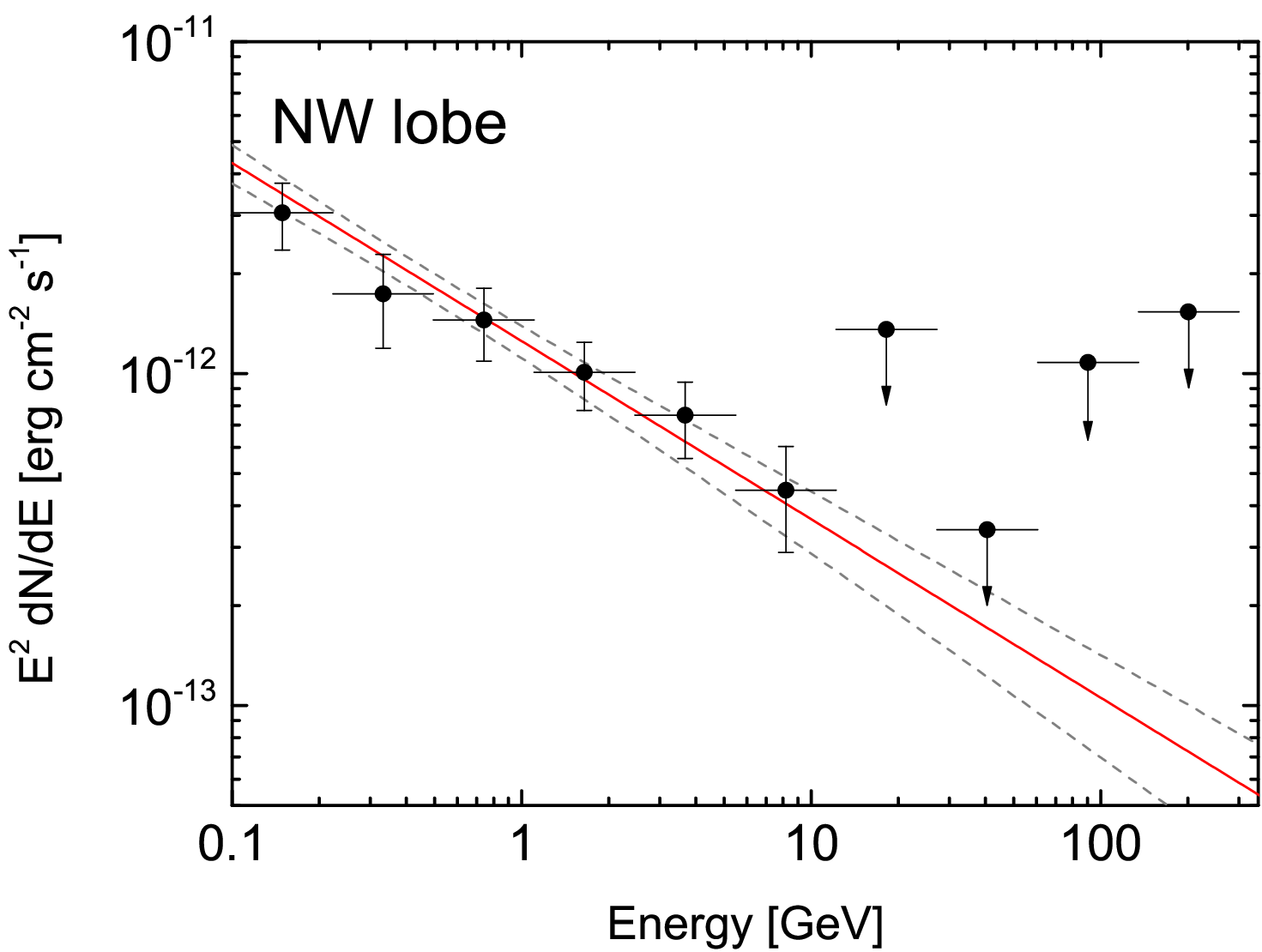}
 \includegraphics[angle=0,scale=0.38]{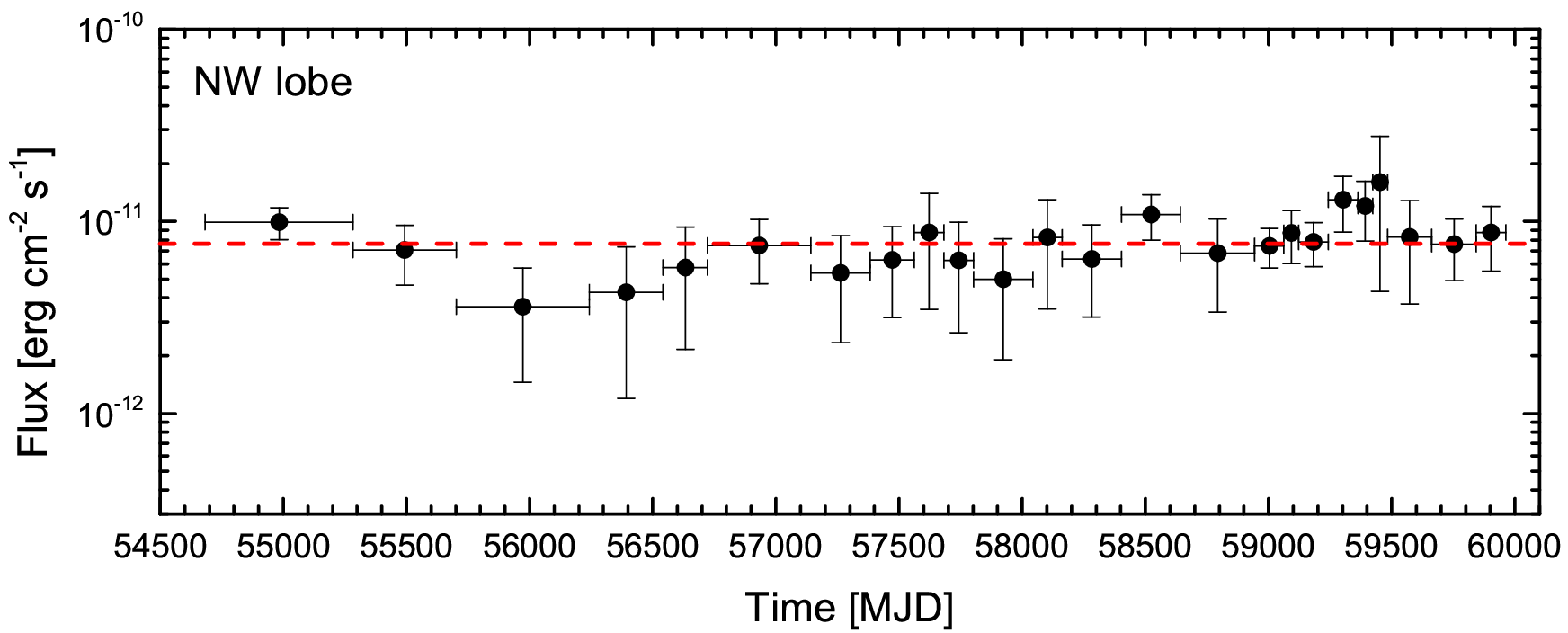}
\caption{Spectra and light curves of core region (the top panels) and NW lobe (the bottom panels) for NGC 6251. They are all derived with the $\sim$15-yr Fermi-LAT observation data in the 0.1--300 GeV band. The solid red lines and gray dashed lines (in two left panels) depict the fitted spectra along with their respective 1$\sigma$ uncertainties. If TS$<$4, an upper limit is presented for that energy bin. The light curves are derived with an adaptive-binning method based on a criterion of TS$\ge$9 for each time bin, where the minimum time-bin step is 60 days. The data points of each time bin in the two light curves are simultaneously obtained. The horizontal red dashed lines represent the $\sim$15-yr average flux for each region, i.e., $F_{\rm 0.1-300~GeV}=(4.86\pm0.61)\times10^{-12}$~erg~cm$^{-2}$~s$^{-1}$ for core region and $F_{\rm 0.1-300~GeV}=(7.67\pm0.76)\times10^{-12}$~erg~cm$^{-2}$~s$^{-1}$ for NW lobe.}\label{extended}
\end{figure}

Through the RMT model analysis, we obtain the spectra and long-term light curves of lobes and core region, as shown in Figure \ref{extended}. The spectrum of the core region still requires a LP function to fit with $\Gamma_{\gamma}=1.80\pm0.13$ and $\beta=0.28\pm0.06$, which is more curved and harder than the total integrated-spectrum of 4FGL J1630.6+8234 (Figure \ref{point}). The $\sim$15-yr average flux of the core region in the 0.1--300 GeV band is $F_{\rm 0.1-300~GeV}=(4.86\pm0.61)\times10^{-12}$~erg~cm$^{-2}$~s$^{-1}$ with TS$\sim$508.55. The spectra of two lobes can be well explained with a PL form. We obtain $\Gamma_{\gamma}=2.54\pm0.06$ and an average flux of $F_{\rm 0.1-300~GeV}=(7.67\pm0.76)\times10^{-12}$~erg~cm$^{-2}$~s$^{-1}$ with TS$\sim$600.31 for the NW lobe. The SE lobe displays a very steep spectrum in the Fermi-LAT energy band with $\Gamma_{\gamma}=7.01\pm0.19$. The average flux $F_{\rm 0.1-300~GeV}=(0.96\pm0.23)\times10^{-12}$~erg~cm$^{-2}$~s$^{-1}$ and the TS value 30.96 are far lower than that of core region and NW lobe. The results are shown in Table \ref{tab:model}. Almost no emission above 1 GeV can be inspected in SE lobe, which is consistent with the $\gamma$-ray counts maps as shown in Figure \ref{countsmap4} in the Appendix. The long-term light curves of core region and NW lobe under the case of the RMT model are simultaneously produced by fixing the parameters of SE lobe and other sources. The adaptive-binning method is also used and a minimum time-bin of 60 days is taken. Neither NW lobe nor core region shows significant flux variation of $\gamma$-rays, as illustrated in Figure \ref{extended}.

\subsection{PS Model vs. RMT Model}

Clearly, the TS value and flux of giant NW lobe are higher than those of the core region. The $\gamma$-rays from lobes account for more than 50\% of the total $\gamma$-ray flux of source. We compare the likelihood values of PS model with RMT model and assess the significance by $\rm TS_{ext}=2(log\mathcal{L}_{\rm RMT}-log\mathcal{L}_{\rm PS})$, where the RMT model has four additional free parameters compared to the PS model. We obtain $\rm TS_{ext}=78.28$, indicating that the RMT model is statistically favored over the PS model at a confidence level of 8.1$\sigma$. 

As described above, the absence of significant variability in the overall long-term light curve of 4FGL J1630.6+8234 (Figure \ref{point}) may be attributed to the dominance of lobe emission; however, no discernible flux variation is observed in the light curve of the core region either (Figure \ref{extended}). In order to further study this issue, we derive the long-term light curve of NGC 6251 in the energy band of 10--300 GeV since the $\gamma$-rays above 10 GeV potentially are dominated by the emission of the core region. Nevertheless, only three detection points can be obtain even using the time bin of 720 days to analyze the observational data, and they do not show the obvious flux variation, as illustrated in Figure \ref{two_lc} in the Appendix. 

\section{SED Constructing and Modeling}
\label{sec3}

The distinct spectral patterns observed in the core region and NW lobe also suggest that $\gamma$-rays from 4FGL J1630.6+8234 have separate origins. We construct the broadband SEDs for the NW lobe and core region, respectively, spanning from the radio to $\gamma$-ray bands, as depicted in Figure \ref{SED}. For the core region, the radio and optical-UV data are collected from \cite{2011A&A...533A..72M}, X-rays are the combined spectrum using the latest 11 XRT observations (see the Appendix), and $\gamma$-rays are obtained by the analysis in this work. For the NW lobe, the radio and X-ray data are collected from \cite{2012ApJ...749...66T} and \cite{2019MNRAS.490.1489P}, while the $\gamma$-ray data are obtained by the analysis in this work. 

We adopt a single-zone leptonic radiation model to fit the broadband SEDs of NW lobe and core region. The model includes synchrotron, SSC (and IC/CMB) processes of the relativistic electrons in radiation regions. The emission region is assumed to be a sphere with a radius $R$, magnetic field strength $B$, the bulk Lorenz factor $\Gamma$ and the Doppler boosting factor $\delta$. The electron distribution is assumed as a broken power-law function, characterized by an electron density parameter $N_0$, a break energy $\gamma_{\rm b}$, and indices $p_1$ and $p_2$ in the range of [$\gamma_{\min}$, $\gamma_{\max}$]. The synchrotron-self-absorption, Klein--Nishina effect, and the extragalactic background light (EBL) absorption \citep{2008A&A...487..837F} are also taken into account during the SED modeling. 

\subsection{Core Region}

We only consider the synchrotron and SSC processes to reproduce the observed SED of core region, same model as the previous works for NGC 6251 \citep{2003ApJ...597..166C,2011A&A...533A..72M,2017RAA....17...90X}. The size of the radiation region is determined as $R=\delta c \Delta t/(1+z)$, where $\Delta t$ is set to 10 days and $c$ represents the speed of light. The values of $p_1$ and $p_2$ are obtained based on the spectral indices at the radio/X-ray and $\gamma$-ray bands, respectively. $\gamma_{\min}$ cannot be constrained and is set as 200 while $\gamma_{\max}$ is roughly constrained by the last data point in the $\gamma$-ray band. We adjust the parameters $B$, $\delta$, $\gamma_{\rm b}$, and $N_0$ to represent the observed SED of core region. It should be noted that these parameter values may not be unique, in particular, $B$ and $\delta$ are degenerate \citep{2012ApJ...752..157Z}. The fitting parameters are given in Table \ref{tab:SED}. 

The model can well reproduce the observed SED of the core region, as illustrated in Figure \ref{SED}. We obtain $\delta=3.4$, which is lower than the typical values observed in blazars, coinciding with the uniform model of RL-AGNs \citep{1995PASP..107..803U}. The same model has been previously employed to explain the archival SEDs of the core region for NGC 6251, where $B=0.03$ G and $\delta=3.2$ were reported \citep{2003ApJ...597..166C}, as well as $B=0.037$ G and $\delta=2.4$ \citep{2011A&A...533A..72M}. Our findings of $B=0.04$ G and $\delta$=3.4 are roughly consistent with theirs.

\subsection{NW Lobe}

We employ the synchrotron+SSC+IC/CMB processes to fit the broadband SED of NW lobe. Considering no relativistic motion for the giant lobe, $\delta=\Gamma=1$ is adopted. In the comoving frame, the CMB energy density is given by $U_{\rm CMB}^{'} = \frac{4}{3} \Gamma^{2} (1+z)^{4} U_{\rm CMB}$ \citep{1994ApJS...90..945D,2006ApJ...653L...5G}, where $U_{\rm CMB} = 4.2 \times 10^{-13}$ erg cm$^{-3}$. The radius of the radiation region is taken as $R=185$ kpc \citep{2012ApJ...749...66T}. The values of $p_1$ and $p_2$ are determined from the spectral indices at radio and $\gamma$-ray bands, respectively. $\gamma_{\min}$ cannot be constrained and is set as 100 while $\gamma_{\max}$ is roughly constrained by the last data point in the $\gamma$-ray band. We adjust the values of parameters $B$, $\gamma_{\rm b}$, and $N_0$ to reproduce the observed data. The fitting parameters are provided in Table \ref{tab:SED}. 

\begin{figure}
 \centering
    \includegraphics[angle=0,scale=0.35]{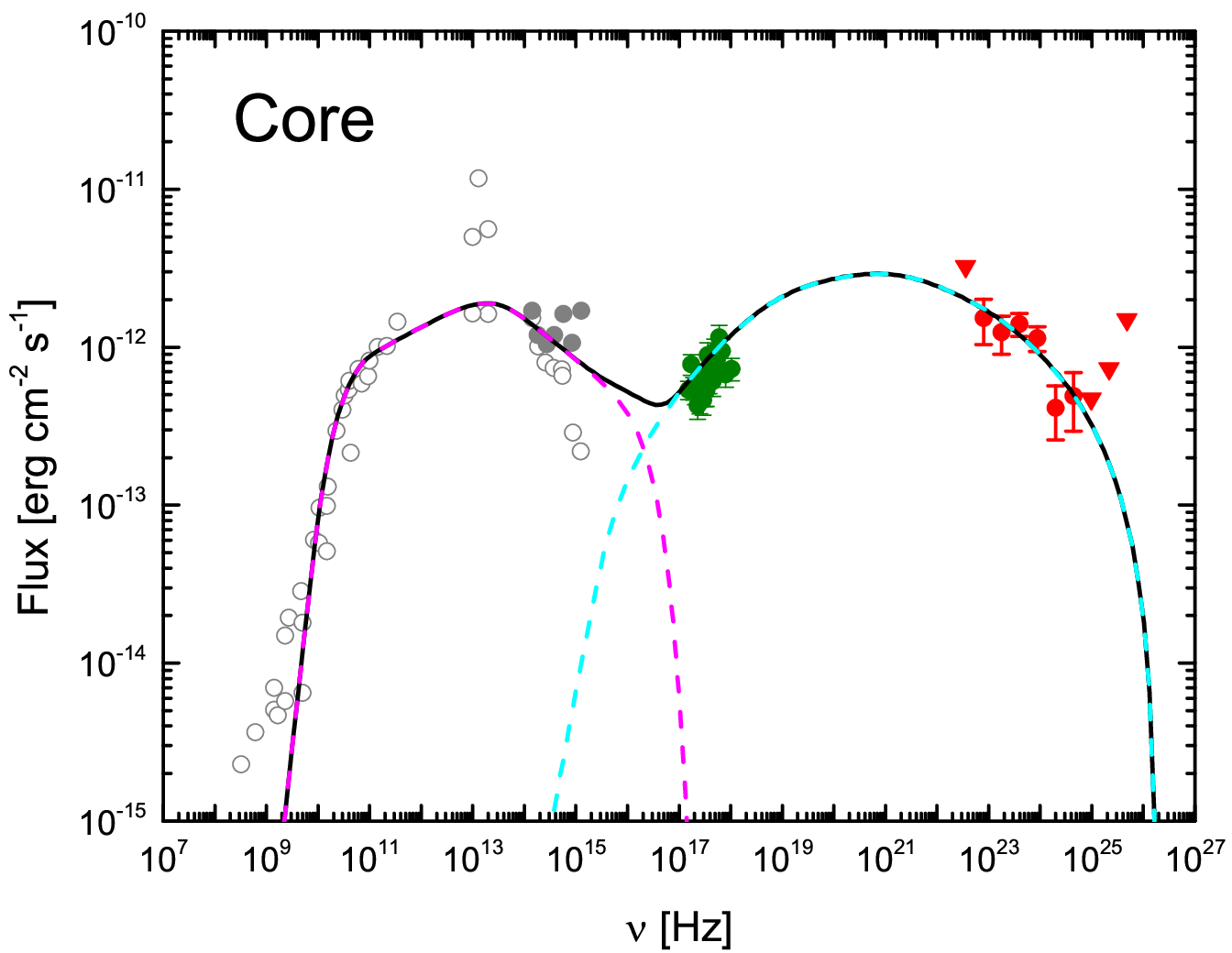}
   \includegraphics[angle=0,scale=0.35]{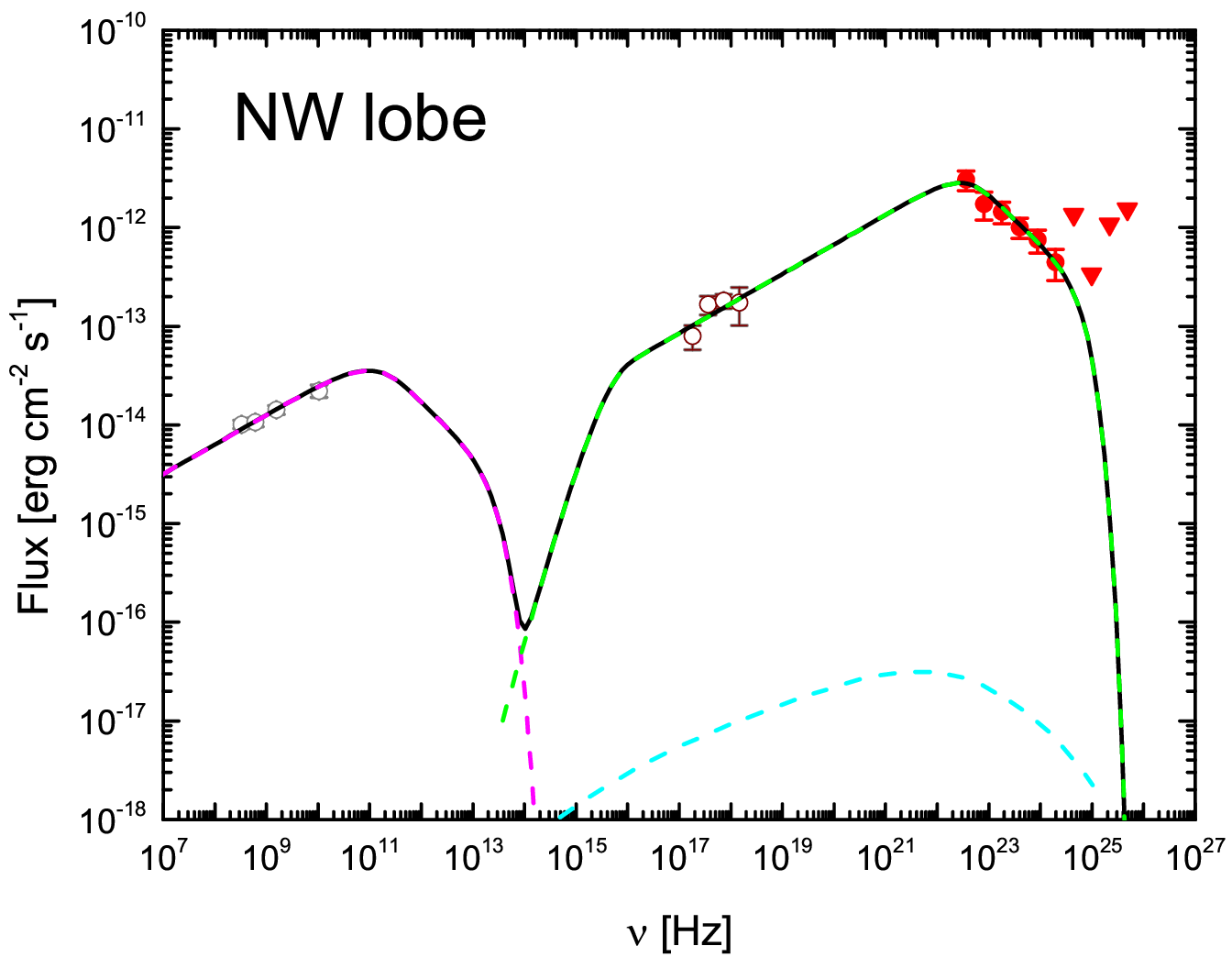}
\caption{Broadband SEDs of the core region (left panel) and NW lobe (right panel) along with the fitting results. The magenta, cyan, and green dashed lines represent the synchrotron, SSC and IC/CMB components, respectively. The black solid lines show the sum of each emission component. For the core region, the data from radio to optical-UV (grey circles) are taken from \cite{2011A&A...533A..72M}, where the gray solid circles indicate the data corrected the dust extinction. The olive circles are the Swift-XRT data obtained in this work. The red symbols are the Fermi-LAT data and are same as that in Figure \ref{extended}. For the NW lobe, the radio (gray opened circles) and X-ray (brown opened circles) data are taken from \cite{2012ApJ...749...66T} and \cite{2019MNRAS.490.1489P}. The red symbols are from the analysis in this work and taken from Figure \ref{extended}. }\label{SED}
\end{figure}

As depicted in Figure \ref{SED}, the emission from the X-ray to $\gamma$-ray bands is dominantly produced by the IC/CMB process. In comparison to the IC/CMB component, the contribution from SSC is negligible. The previous study utilized the IC/(CMB+EBL) model, employing a double broken power-law electron distribution to represent the X-ray data of the NW lobe in NGC 6251 \citep{2012ApJ...749...66T}, and they yielded $B=0.37~\mu$G and $\gamma_{\max}=10^6$ (corresponding to the value of $\gamma_{\rm b}$ in our work). Using a similar model with a power-law electron distribution to fit the SED, \cite{2019MNRAS.490.1489P} obtained $B=0.40~\mu$G, $\gamma_{\max}=1.1\times10^6$, and the slope index of $q=2.5$. Our results are consistent with theirs.

\section{Discussion and Conclusions} 
\label{sec4}

RG NGC 6251 is reported to be associated with the $\gamma$-ray source 4FGL J1630.6+8234, and 4FGL J1630.6+8234 is identified as a PS in the 4FGL-DR4 \citep{2023arXiv230712546B,2022ApJS..260...53A}. We comprehensively analyze the $\sim$15-yr Fermi-LAT observation data of 4FGL J1630.6+8234 and observe that the giant lobes dominate the detected $\gamma$-rays of NGC 6251. By comparing the maximum likelihood values between the RMT model analysis and a PS model analysis, we find that the RMT model analysis is more preferable to describe the $\gamma$-ray emission of NGC 6251 at a confidence level of 8.1$\sigma$.

The spectrum of the core region is better described by a LP function compared to a PL function, while the spectra of the two lobes can be well explained with a PL form. The distinct spectral shapes between core region and lobes also demonstrate the different origins of $\gamma$-rays. The $\gamma$-ray flux of NW lobe is even higher than that of the core region, to account for more than 50\% of the total $\gamma$-ray flux. But the emission flux from SE lobe is very low, only being $\sim7\%$ of the total flux. The TS value of the SE lobe is only 30.96, while it is TS$\sim$600.31 for the NW lobe and TS$\sim$508.55 for the core region. 

Neither the core nor the NW lobe displays the significant flux variation in their long-term $\gamma$-ray light curves. However, X-ray observations of NGC 6251 using Swift-XRT indeed show flux variations at a confidence level of 4.9$\sigma$ (Figure \ref{two_lc} in the Appendix), which are likely attributed to the core-jet emission. It is worth noting that the core region in the RMT model encompasses the straight jet; the low-level $\gamma$-ray flux variations originating from the radio core might be concealed by the radiation of straight jet. This further supports the conclusion that the dominant contribution to $\gamma$-ray fluxes observed from 4FGL J1630.6+8234 stems from the extended jet substructures of NGC 6251, surpassing that of the radio core. 

On the base of the derived average spectra in X-ray and $\gamma$-ray bands, we construct the broadband SEDs of NW lobe and core region and fit them with an one-zone leptonic model. The $\gamma$-rays of NW lobe together with its X-ray emission are attributed to the IC/CMB by the relativistic electrons while the $\gamma$-rays of core region are produced by SSC process. The derived parameter values of radiation regions are consistent with the previous works for studying NGC 6251. 

Following the examples of Cen A \citep{2010Sci...328..725A} and Fornax A \citep{2016ApJ...826....1A}, NGC 6251 is the third RG with measured extended $\gamma$-ray emission. Our findings would further improve the cognition of the diversity $\gamma$-ray emitters in $\gamma$-ray band. The $\gamma$-rays from the large-scale extended regions of the three RGs can be measured only due to their close proximity and the large angular size. The low spatial resolution of $\gamma$-ray detectors makes it difficult to judge the location of the $\gamma$-ray emission for most sources. On this point, the $\gamma$-ray emission from the large-scale jets of RGs may be universal, as predicted by some theoretic works \citep{2009ApJ...701..423Z,2010ApJ...710.1017Z,2018ApJ...858...27Z}. On the other hand, most detected extragalactic sources in the GeV-TeV $\gamma$-ray band are blazars due to their aligned jets with the strong Doppler amplification of their emission. The lack of the relativistic effect in the large-scale jets implies that the detected $\gamma$-rays from these regions are intrinsically strong. Therefore, the large-scale jets of AGNs may provide more energy input to the intergalactic medium in $\gamma$-ray band \citep{2020Natur.582..356H}.   
  
\acknowledgments

We thank the anonymous referee for the valuable suggestions. This work is supported by the National Natural Science Foundation of China (grants 12022305, 12203022, 11973050).

\bibliography{reference}{}

\begin{thebibliography}{}
\expandafter\ifx\csname natexlab\endcsname\relax\def\natexlab#1{#1}\fi
\providecommand{\url}[1]{\href{#1}{#1}}
\providecommand{\dodoi}[1]{doi:~\href{http://doi.org/#1}{\nolinkurl{#1}}}
\providecommand{\doeprint}[1]{\href{http://ascl.net/#1}{\nolinkurl{http://ascl.net/#1}}}
\providecommand{\doarXiv}[1]{\href{https://arxiv.org/abs/#1}{\nolinkurl{https://arxiv.org/abs/#1}}}

\bibitem[{{Abdo} {et~al.}(2010{\natexlab{a}}){Abdo}, {Ackermann}, {Ajello},
  {Atwood}, {Baldini}, {Ballet}, {Barbiellini}, {Bastieri}, {Baughman},
  {Bechtol}, {Bellazzini}, {Berenji}, {Blandford}, {Bloom}, {Bonamente},
  {Borgland}, {Bregeon}, {Brez}, {Brigida}, {Bruel}, {Burnett}, {Buson},
  {Caliandro}, {Cameron}, {Caraveo}, {Casandjian}, {Cavazzuti}, {Cecchi},
  {{\c{C}}elik}, {Chekhtman}, {Cheung}, {Chiang}, {Ciprini}, {Claus},
  {Cohen-Tanugi}, {Colafrancesco}, {Cominsky}, {Conrad}, {Costamante},
  {Cutini}, {Davis}, {Dermer}, {de Angelis}, {de Palma}, {Digel}, {do Couto e
  Silva}, {Drell}, {Dubois}, {Dumora}, {Farnier}, {Favuzzi}, {Fegan}, {Finke},
  {Focke}, {Fortin}, {Fukazawa}, {Funk}, {Fusco}, {Gargano}, {Gasparrini},
  {Gehrels}, {Georganopoulos}, {Germani}, {Giebels}, {Giglietto}, {Giordano},
  {Giroletti}, {Glanzman}, {Godfrey}, {Grenier}, {Grove}, {Guillemot},
  {Guiriec}, {Hanabata}, {Harding}, {Hayashida}, {Hays}, {Hughes}, {Jackson},
  {J{\'o}hannesson G.}, {Johnson}, {Johnson}, {Johnson}, {Kamae}, {Katagiri},
  {Kataoka}, {Kawai}, {Kerr}, {Kn{\"o}dlseder}, {Kocian}, {Kuss}, {Lande},
  {Latronico}, {Lemoine-Goumard}, {Longo}, {Loparco}, {Lott}, {Lovellette},
  {Lubrano}, {Madejski}, {Makeev}, {Mazziotta}, {McConville}, {McEnery},
  {Meurer}, {Michelson}, {Mitthumsiri}, {Mizuno}, {Moiseev}, {Monte},
  {Monzani}, {Morselli}, {Moskalenko}, {Murgia}, {Nolan}, {Norris}, {Nuss},
  {Ohsugi}, {Omodei}, {Orlando}, {Ormes}, {Paneque}, {Parent}, {Pelassa},
  {Pepe}, {Pesce-Rollins}, {Piron}, {Porter}, {Rain{\`o}}, {Rando}, {Razzano},
  {Razzaque}, {Reimer}, {Reimer}, {Reposeur}, {Ritz}, {Rochester}, {Rodriguez},
  {Romani}, {Roth}, {Ryde}, {Sadrozinski}, {Sambruna}, {Sanchez}, {Sander},
  {Saz Parkinson}, {Scargle}, {Sgr{\`o}}, {Siskind}, {Smith}, {Smith},
  {Spandre}, {Spinelli}, {Starck}, {Stawarz}, {Strickman}, {Suson}, {Tajima},
  {Takahashi}, {Takahashi}, {Tanaka}, {Thayer}, {Thayer}, {Thompson},
  {Tibaldo}, {Torres}, {Tosti}, {Tramacere}, {Uchiyama}, {Vasileiou},
  {Vilchez}, {Vitale}, {Waite}, {Wallace}, {Wang}, {Winer}, {Wood}, {Ylinen},
  {Ziegler}, {Hardcastle}, {Kazanas}, \& {Fermi LAT
  Collaboration}}]{2010Sci...328..725A}
{Abdo}, A.~A., {Ackermann}, M., {Ajello}, M., {et~al.} 2010{\natexlab{a}},
  Science, 328, 725, \dodoi{10.1126/science.1184656}

\bibitem[{{Abdo} {et~al.}(2010{\natexlab{b}}){Abdo}, {Ackermann}, {Ajello},
  {Allafort}, {Antolini}, {Atwood}, {Axelsson}, {Baldini}, {Ballet},
  {Barbiellini}, {Bastieri}, {Baughman}, {Bechtol}, {Bellazzini}, {Belli},
  {Berenji}, {Bisello}, {Blandford}, {Bloom}, {Bonamente}, {Bonnell},
  {Borgland}, {Bouvier}, {Bregeon}, {Brez}, {Brigida}, {Bruel}, {Burnett},
  {Busetto}, {Buson}, {Caliandro}, {Cameron}, {Campana}, {Canadas}, {Caraveo},
  {Carrigan}, {Casandjian}, {Cavazzuti}, {Ceccanti}, {Cecchi}, {{\c{C}}elik},
  {Charles}, {Chekhtman}, {Cheung}, {Chiang}, {Cillis}, {Ciprini}, {Claus},
  {Cohen-Tanugi}, {Conrad}, {Corbet}, {Davis}, {DeKlotz}, {den Hartog},
  {Dermer}, {de Angelis}, {de Luca}, {de Palma}, {Digel}, {Dormody}, {Silva},
  {Drell}, {Dubois}, {Dumora}, {Fabiani}, {Farnier}, {Favuzzi}, {Fegan},
  {Ferrara}, {Focke}, {Fortin}, {Frailis}, {Fukazawa}, {Funk}, {Fusco},
  {Gargano}, {Gasparrini}, {Gehrels}, {Germani}, {Giavitto}, {Giebels},
  {Giglietto}, {Giommi}, {Giordano}, {Giroletti}, {Glanzman}, {Godfrey},
  {Grenier}, {Grondin}, {Grove}, {Guillemot}, {Guiriec}, {Gustafsson},
  {Hadasch}, {Hanabata}, {Harding}, {Hayashida}, {Hays}, {Healey}, {Hill},
  {Horan}, {Hughes}, {Iafrate}, {J{\'o}hannesson}, {Johnson}, {Johnson},
  {Johnson}, {Johnson}, {Kamae}, {Katagiri}, {Kataoka}, {Kawai}, {Kerr},
  {Kn{\"o}dlseder}, {Kocevski}, {Kuss}, {Lande}, {Landriu}, {Latronico}, {Lee},
  {Lemoine-Goumard}, {Lionetto}, {Llena Garde}, {Longo}, {Loparco}, {Lott},
  {Lovellette}, {Lubrano}, {Madejski}, {Makeev}, {Marangelli}, {Marelli},
  {Massaro}, {Mazziotta}, {McConville}, {McEnery}, {Michelson}, {Minuti},
  {Mitthumsiri}, {Mizuno}, {Moiseev}, {Mongelli}, {Monte}, {Monzani},
  {Moretti}, {Morselli}, {Moskalenko}, {Murgia}, {Nakajima}, {Nakamori},
  {Naumann-Godo}, {Nolan}, {Norris}, {Nuss}, {Ohno}, {Ohsugi}, {Omodei},
  {Orlando}, {Ormes}, {Ozaki}, {Paccagnella}, {Paneque}, {Panetta}, {Parent},
  {Pelassa}, {Pepe}, {Pesce-Rollins}, {Pinchera}, {Piron}, {Porter}, {Poupard},
  {Rain{\`o}}, {Rando}, {Ray}, {Razzano}, {Razzaque}, {Rea}, {Reimer},
  {Reimer}, {Reposeur}, {Ripken}, {Ritz}, {Rochester}, {Rodriguez}, {Romani},
  {Roth}, {Sadrozinski}, {Salvetti}, {Sanchez}, {Sander}, {Saz Parkinson},
  {Scargle}, {Schalk}, {Scolieri}, {Sgr{\`o}}, {Shaw}, {Siskind}, {Smith},
  {Smith}, {Spandre}, {Spinelli}, {Starck}, {Stephens}, {Striani}, {Strickman},
  {Strong}, {Suson}, {Tajima}, {Takahashi}, {Takahashi}, {Tanaka}, {Thayer},
  {Thayer}, {Thompson}, {Tibaldo}, {Tibolla}, {Tinebra}, {Torres}, {Tosti},
  {Tramacere}, {Uchiyama}, {Usher}, {Van Etten}, {Vasileiou}, {Vilchez},
  {Vitale}, {Waite}, {Wallace}, {Wang}, {Watters}, {Winer}, {Wood}, {Yang},
  {Ylinen}, {Ziegler}, \& {Fermi LAT Collaboration}}]{2010ApJS..188..405A}
---. 2010{\natexlab{b}}, \apjs, 188, 405, \dodoi{10.1088/0067-0049/188/2/405}

\bibitem[{{Abdo} {et~al.}(2010{\natexlab{c}}){Abdo}, {Ackermann}, {Ajello},
  {Baldini}, {Ballet}, {Barbiellini}, {Bastieri}, {Bechtol}, {Bellazzini},
  {Berenji}, {Blandford}, {Bloom}, {Bonamente}, {Borgland}, {Bouvier},
  {Brandt}, {Bregeon}, {Brez}, {Brigida}, {Bruel}, {Buehler}, {Burnett},
  {Buson}, {Caliandro}, {Cameron}, {Cannon}, {Caraveo}, {Carrigan},
  {Casandjian}, {Cavazzuti}, {Cecchi}, {{\c{C}}elik}, {Celotti}, {Charles},
  {Chekhtman}, {Chen}, {Cheung}, {Chiang}, {Ciprini}, {Claus}, {Cohen-Tanugi},
  {Colafrancesco}, {Conrad}, {Davis}, {Dermer}, {de Angelis}, {de Palma},
  {Silva}, {Drell}, {Dubois}, {Favuzzi}, {Fegan}, {Ferrara}, {Fortin},
  {Frailis}, {Fukazawa}, {Fusco}, {Gargano}, {Gasparrini}, {Gehrels},
  {Germani}, {Giglietto}, {Giommi}, {Giordano}, {Giroletti}, {Glanzman},
  {Godfrey}, {Grandi}, {Grenier}, {Grove}, {Guillemot}, {Guiriec}, {Hadasch},
  {Hayashida}, {Hays}, {Horan}, {Hughes}, {Jackson}, {J{\'o}hannesson},
  {Johnson}, {Johnson}, {Kamae}, {Katagiri}, {Kataoka}, {Kn{\"o}dlseder},
  {Kuss}, {Lande}, {Latronico}, {Lee}, {Lemoine-Goumard}, {Llena Garde},
  {Longo}, {Loparco}, {Lott}, {Lovellette}, {Lubrano}, {Madejski}, {Makeev},
  {Malaguti}, {Mazziotta}, {McConville}, {McEnery}, {Michelson}, {Migliori},
  {Mitthumsiri}, {Mizuno}, {Monte}, {Monzani}, {Morselli}, {Moskalenko},
  {Murgia}, {Naumann-Godo}, {Nestoras}, {Nolan}, {Norris}, {Nuss}, {Ohsugi},
  {Okumura}, {Omodei}, {Orlando}, {Ormes}, {Paneque}, {Panetta}, {Parent},
  {Pelassa}, {Pepe}, {Persic}, {Pesce-Rollins}, {Piron}, {Porter}, {Rain{\`o}},
  {Rando}, {Razzano}, {Razzaque}, {Reimer}, {Reimer}, {Reyes}, {Roth},
  {Sadrozinski}, {Sanchez}, {Sander}, {Scargle}, {Sgr{\`o}}, {Siskind},
  {Smith}, {Spandre}, {Spinelli}, {Stawarz}, {Stecker}, {Strickman}, {Suson},
  {Takahashi}, {Tanaka}, {Thayer}, {Thayer}, {Thompson}, {Tibaldo}, {Torres},
  {Torresi}, {Tosti}, {Tramacere}, {Uchiyama}, {Usher}, {Vandenbroucke},
  {Vasileiou}, {Vilchez}, {Villata}, {Vitale}, {Waite}, {Wang}, {Winer},
  {Wood}, {Yang}, {Ylinen}, \& {Ziegler}}]{2010ApJ...720..912A}
---. 2010{\natexlab{c}}, \apj, 720, 912, \dodoi{10.1088/0004-637X/720/1/912}

\bibitem[{{Abdollahi} {et~al.}(2022){Abdollahi}, {Acero}, {Baldini}, {Ballet},
  {Bastieri}, {Bellazzini}, {Berenji}, {Berretta}, {Bissaldi}, {Blandford},
  {Bloom}, {Bonino}, {Brill}, {Britto}, {Bruel}, {Burnett}, {Buson}, {Cameron},
  {Caputo}, {Caraveo}, {Castro}, {Chaty}, {Cheung}, {Chiaro}, {Cibrario},
  {Ciprini}, {Coronado-Bl{\'a}zquez}, {Crnogorcevic}, {Cutini}, {D'Ammando},
  {De Gaetano}, {Digel}, {Di Lalla}, {Dirirsa}, {Di Venere}, {Dom{\'\i}nguez},
  {Fallah Ramazani}, {Fegan}, {Ferrara}, {Fiori}, {Fleischhack}, {Franckowiak},
  {Fukazawa}, {Funk}, {Fusco}, {Galanti}, {Gammaldi}, {Gargano}, {Garrappa},
  {Gasparrini}, {Giacchino}, {Giglietto}, {Giordano}, {Giroletti}, {Glanzman},
  {Green}, {Grenier}, {Grondin}, {Guillemot}, {Guiriec}, {Gustafsson},
  {Harding}, {Hays}, {Hewitt}, {Horan}, {Hou}, {J{\'o}hannesson}, {Karwin},
  {Kayanoki}, {Kerr}, {Kuss}, {Landriu}, {Larsson}, {Latronico},
  {Lemoine-Goumard}, {Li}, {Liodakis}, {Longo}, {Loparco}, {Lott}, {Lubrano},
  {Maldera}, {Malyshev}, {Manfreda}, {Mart{\'\i}-Devesa}, {Mazziotta}, {Mereu},
  {Meyer}, {Michelson}, {Mirabal}, {Mitthumsiri}, {Mizuno}, {Moiseev},
  {Monzani}, {Morselli}, {Moskalenko}, {Negro}, {Nuss}, {Omodei}, {Orienti},
  {Orlando}, {Paneque}, {Pei}, {Perkins}, {Persic}, {Pesce-Rollins},
  {Petrosian}, {Pillera}, {Poon}, {Porter}, {Principe}, {Rain{\`o}}, {Rando},
  {Rani}, {Razzano}, {Razzaque}, {Reimer}, {Reimer}, {Reposeur},
  {S{\'a}nchez-Conde}, {Saz Parkinson}, {Scotton}, {Serini}, {Sgr{\`o}},
  {Siskind}, {Smith}, {Spandre}, {Spinelli}, {Sueoka}, {Suson}, {Tajima},
  {Tak}, {Thayer}, {Thompson}, {Torres}, {Troja}, {Valverde}, {Wood}, \&
  {Zaharijas}}]{2022ApJS..260...53A}
{Abdollahi}, S., {Acero}, F., {Baldini}, L., {et~al.} 2022, \apjs, 260, 53,
  \dodoi{10.3847/1538-4365/ac6751}

\bibitem[{{Acero} {et~al.}(2015){Acero}, {Ackermann}, {Ajello}, {Albert},
  {Atwood}, {Axelsson}, {Baldini}, {Ballet}, {Barbiellini}, {Bastieri},
  {Belfiore}, {Bellazzini}, {Bissaldi}, {Blandford}, {Bloom}, {Bogart},
  {Bonino}, {Bottacini}, {Bregeon}, {Britto}, {Bruel}, {Buehler}, {Burnett},
  {Buson}, {Caliandro}, {Cameron}, {Caputo}, {Caragiulo}, {Caraveo},
  {Casandjian}, {Cavazzuti}, {Charles}, {Chaves}, {Chekhtman}, {Cheung},
  {Chiang}, {Chiaro}, {Ciprini}, {Claus}, {Cohen-Tanugi}, {Cominsky}, {Conrad},
  {Cutini}, {D'Ammando}, {de Angelis}, {DeKlotz}, {de Palma}, {Desiante},
  {Digel}, {Di Venere}, {Drell}, {Dubois}, {Dumora}, {Favuzzi}, {Fegan},
  {Ferrara}, {Finke}, {Franckowiak}, {Fukazawa}, {Funk}, {Fusco}, {Gargano},
  {Gasparrini}, {Giebels}, {Giglietto}, {Giommi}, {Giordano}, {Giroletti},
  {Glanzman}, {Godfrey}, {Grenier}, {Grondin}, {Grove}, {Guillemot}, {Guiriec},
  {Hadasch}, {Harding}, {Hays}, {Hewitt}, {Hill}, {Horan}, {Iafrate}, {Jogler},
  {J{\'o}hannesson}, {Johnson}, {Johnson}, {Johnson}, {Johnson}, {Kamae},
  {Kataoka}, {Katsuta}, {Kuss}, {La Mura}, {Landriu}, {Larsson}, {Latronico},
  {Lemoine-Goumard}, {Li}, {Li}, {Longo}, {Loparco}, {Lott}, {Lovellette},
  {Lubrano}, {Madejski}, {Massaro}, {Mayer}, {Mazziotta}, {McEnery},
  {Michelson}, {Mirabal}, {Mizuno}, {Moiseev}, {Mongelli}, {Monzani},
  {Morselli}, {Moskalenko}, {Murgia}, {Nuss}, {Ohno}, {Ohsugi}, {Omodei},
  {Orienti}, {Orlando}, {Ormes}, {Paneque}, {Panetta}, {Perkins},
  {Pesce-Rollins}, {Piron}, {Pivato}, {Porter}, {Racusin}, {Rando}, {Razzano},
  {Razzaque}, {Reimer}, {Reimer}, {Reposeur}, {Rochester}, {Romani},
  {Salvetti}, {S{\'a}nchez-Conde}, {Saz Parkinson}, {Schulz}, {Siskind},
  {Smith}, {Spada}, {Spandre}, {Spinelli}, {Stephens}, {Strong}, {Suson},
  {Takahashi}, {Takahashi}, {Tanaka}, {Thayer}, {Thayer}, {Thompson},
  {Tibaldo}, {Tibolla}, {Torres}, {Torresi}, {Tosti}, {Troja}, {Van Klaveren},
  {Vianello}, {Winer}, {Wood}, {Wood}, {Zimmer}, \& {Fermi-LAT
  Collaboration}}]{2015ApJS..218...23A}
{Acero}, F., {Ackermann}, M., {Ajello}, M., {et~al.} 2015, \apjs, 218, 23,
  \dodoi{10.1088/0067-0049/218/2/23}

\bibitem[{{Ackermann} {et~al.}(2016){Ackermann}, {Ajello}, {Baldini}, {Ballet},
  {Barbiellini}, {Bastieri}, {Bellazzini}, {Bissaldi}, {Blandford}, {Bloom},
  {Bonino}, {Brandt}, {Bregeon}, {Bruel}, {Buehler}, {Buson}, {Caliandro},
  {Cameron}, {Caragiulo}, {Caraveo}, {Cavazzuti}, {Cecchi}, {Charles},
  {Chekhtman}, {Cheung}, {Chiaro}, {Ciprini}, {Cohen}, {Cohen-Tanugi},
  {Costanza}, {Cutini}, {D'Ammando}, {Davis}, {de Angelis}, {de Palma},
  {Desiante}, {Digel}, {Di Lalla}, {Di Mauro}, {Di Venere}, {Favuzzi}, {Fegan},
  {Ferrara}, {Focke}, {Fukazawa}, {Funk}, {Fusco}, {Gargano}, {Gasparrini},
  {Georganopoulos}, {Giglietto}, {Giordano}, {Giroletti}, {Godfrey}, {Green},
  {Grenier}, {Guiriec}, {Hays}, {Hewitt}, {Hill}, {Jogler}, {J{\'o}hannesson},
  {Kensei}, {Kuss}, {Larsson}, {Latronico}, {Li}, {Li}, {Longo}, {Loparco},
  {Lubrano}, {Magill}, {Maldera}, {Manfreda}, {Mayer}, {Mazziotta},
  {McConville}, {McEnery}, {Michelson}, {Mitthumsiri}, {Mizuno}, {Monzani},
  {Morselli}, {Moskalenko}, {Murgia}, {Negro}, {Nuss}, {Ohno}, {Ohsugi},
  {Orienti}, {Orlando}, {Ormes}, {Paneque}, {Perkins}, {Pesce-Rollins},
  {Piron}, {Pivato}, {Porter}, {Rain{\`o}}, {Rando}, {Razzano}, {Reimer},
  {Reimer}, {Schmid}, {Sgr{\`o}}, {Simone}, {Siskind}, {Spada}, {Spandre},
  {Spinelli}, {Stawarz}, {Takahashi}, {Thayer}, {Thompson}, {Torres}, {Tosti},
  {Troja}, {Vianello}, {Wood}, {Wood}, {Zimmer}, \& {Fermi LAT
  Collaboration}}]{2016ApJ...826....1A}
{Ackermann}, M., {Ajello}, M., {Baldini}, L., {et~al.} 2016, \apj, 826, 1,
  \dodoi{10.3847/0004-637X/826/1/1}

\bibitem[{{Ballet} {et~al.}(2023){Ballet}, {Bruel}, {Burnett}, {Lott}, \& {The
  Fermi-LAT collaboration}}]{2023arXiv230712546B}
{Ballet}, J., {Bruel}, P., {Burnett}, T.~H., {Lott}, B., \& {The Fermi-LAT
  collaboration}. 2023, arXiv e-prints, arXiv:2307.12546,
  \dodoi{10.48550/arXiv.2307.12546}

\bibitem[{{Birkinshaw} \& {Worrall}(1993)}]{1993ApJ...412..568B}
{Birkinshaw}, M., \& {Worrall}, D.~M. 1993, \apj, 412, 568,
  \dodoi{10.1086/172944}

\bibitem[{{Burrows} {et~al.}(2005){Burrows}, {Hill}, {Nousek}, {Kennea},
  {Wells}, {Osborne}, {Abbey}, {Beardmore}, {Mukerjee}, {Short}, {Chincarini},
  {Campana}, {Citterio}, {Moretti}, {Pagani}, {Tagliaferri}, {Giommi},
  {Capalbi}, {Tamburelli}, {Angelini}, {Cusumano}, {Br{\"a}uninger}, {Burkert},
  \& {Hartner}}]{2005SSRv..120..165B}
{Burrows}, D.~N., {Hill}, J.~E., {Nousek}, J.~A., {et~al.} 2005, \ssr, 120,
  165, \dodoi{10.1007/s11214-005-5097-2}

\bibitem[{{Cantwell} {et~al.}(2020){Cantwell}, {Bray}, {Croston}, {Scaife},
  {Mulcahy}, {Best}, {Br{\"u}ggen}, {Brunetti}, {Callingham}, {Clarke},
  {Hardcastle}, {Harwood}, {Heald}, {Heesen}, {Iacobelli}, {Jamrozy},
  {Morganti}, {Orr{\'u}}, {O'Sullivan}, {Riseley}, {R{\"o}ttgering},
  {Shulevski}, {Sridhar}, {Tasse}, \& {Van Eck}}]{2020MNRAS.495..143C}
{Cantwell}, T.~M., {Bray}, J.~D., {Croston}, J.~H., {et~al.} 2020, \mnras, 495,
  143, \dodoi{10.1093/mnras/staa1160}

\bibitem[{{Chen} {et~al.}(2022){Chen}, {Laor}, \&
  {Behar}}]{2022MNRAS.515.1723C}
{Chen}, S., {Laor}, A., \& {Behar}, E. 2022, \mnras, 515, 1723,
  \dodoi{10.1093/mnras/stac1891}

\bibitem[{{Chiaberge} {et~al.}(2003){Chiaberge}, {Gilli}, {Capetti}, \&
  {Macchetto}}]{2003ApJ...597..166C}
{Chiaberge}, M., {Gilli}, R., {Capetti}, A., \& {Macchetto}, F.~D. 2003, \apj,
  597, 166, \dodoi{10.1086/378289}

\bibitem[{{Dermer} \& {Schlickeiser}(1994)}]{1994ApJS...90..945D}
{Dermer}, C.~D., \& {Schlickeiser}, R. 1994, \apjs, 90, 945,
  \dodoi{10.1086/191929}

\bibitem[{{Evans} {et~al.}(2005){Evans}, {Hardcastle}, {Croston}, {Worrall}, \&
  {Birkinshaw}}]{2005MNRAS.359..363E}
{Evans}, D.~A., {Hardcastle}, M.~J., {Croston}, J.~H., {Worrall}, D.~M., \&
  {Birkinshaw}, M. 2005, \mnras, 359, 363,
  \dodoi{10.1111/j.1365-2966.2005.08900.x}

\bibitem[{{Fermi Science Support Development Team}(2019)}]{2019ascl.soft05011F}
{Fermi Science Support Development Team}. 2019, {Fermitools: Fermi Science
  Tools}, Astrophysics Source Code Library, record ascl:1905.011.
\newblock \doeprint{1905.011}

\bibitem[{{Foschini} {et~al.}(2005){Foschini}, {Chiaberge}, {Grandi},
  {Grenier}, {Guainazzi}, {Hermsen}, {Palumbo}, {Rodriguez}, {Chaty}, {Corbel},
  {Di Cocco}, {Kuiper}, \& {Malaguti}}]{2005A&A...433..515F}
{Foschini}, L., {Chiaberge}, M., {Grandi}, P., {et~al.} 2005, \aap, 433, 515,
  \dodoi{10.1051/0004-6361:20042353}

\bibitem[{{Foschini} {et~al.}(2022){Foschini}, {Lister}, {Andernach}, {Ciroi},
  {Marziani}, {Ant{\'o}n}, {Berton}, {Dalla Bont{\`a}}, {J{\"a}rvel{\"a}},
  {March{\~a}}, {Romano}, {Tornikoski}, {Vercellone}, \&
  {Vietri}}]{2022Univ....8..587F}
{Foschini}, L., {Lister}, M.~L., {Andernach}, H., {et~al.} 2022, Universe, 8,
  587, \dodoi{10.3390/universe8110587}

\bibitem[{{Franceschini} {et~al.}(2008){Franceschini}, {Rodighiero}, \&
  {Vaccari}}]{2008A&A...487..837F}
{Franceschini}, A., {Rodighiero}, G., \& {Vaccari}, M. 2008, \aap, 487, 837,
  \dodoi{10.1051/0004-6361:200809691}

\bibitem[{{Fukazawa} {et~al.}(2015){Fukazawa}, {Finke}, {Stawarz}, {Tanaka},
  {Itoh}, \& {Tokuda}}]{2015ApJ...798...74F}
{Fukazawa}, Y., {Finke}, J., {Stawarz}, {\L}., {et~al.} 2015, \apj, 798, 74,
  \dodoi{10.1088/0004-637X/798/2/74}

\bibitem[{{Gehrels} \& {Swift}(2004)}]{2004AAS...20511601G}
{Gehrels}, N., \& {Swift}. 2004, in American Astronomical Society Meeting
  Abstracts, Vol. 205, American Astronomical Society Meeting Abstracts, 116.01

\bibitem[{{Georganopoulos} {et~al.}(2006){Georganopoulos}, {Perlman},
  {Kazanas}, \& {McEnery}}]{2006ApJ...653L...5G}
{Georganopoulos}, M., {Perlman}, E.~S., {Kazanas}, D., \& {McEnery}, J. 2006,
  \apjl, 653, L5, \dodoi{10.1086/510452}

\bibitem[{{H.~E.~S.~S. Collaboration} {et~al.}(2020){H.~E.~S.~S.
  Collaboration}, {Abdalla}, {Adam}, {Aharonian}, {Ait Benkhali},
  {Ang{\"u}ner}, {Arakawa}, {Arcaro}, {Armand}, {Ashkar}, {Backes}, {Barbosa
  Martins}, {Barnard}, {Becherini}, {Berge}, {Bernl{\"o}hr}, {Blackwell},
  {B{\"o}ttcher}, {Boisson}, {Bolmont}, {Bonnefoy}, {Bregeon}, {Breuhaus},
  {Brun}, {Brun}, {Bryan}, {B{\"u}chele}, {Bulik}, {Bylund}, {Capasso},
  {Caroff}, {Carosi}, {Casanova}, {Cerruti}, {Chand}, {Chandra}, {Chen},
  {Colafrancesco}, {Cury{\l}o}, {Davids}, {Deil}, {Devin}, {deWilt}, {Dirson},
  {Djannati-Ata{\"\i}}, {Dmytriiev}, {Donath}, {Doroshenko}, {Drury}, {Dyks},
  {Egberts}, {Emery}, {Ernenwein}, {Eschbach}, {Feijen}, {Fegan}, {Fiasson},
  {Fontaine}, {Funk}, {F{\"u}{\ss}ling}, {Gabici}, {Gallant}, {Gat{\'e}},
  {Giavitto}, {Glawion}, {Glicenstein}, {Gottschall}, {Grondin}, {Hahn},
  {Haupt}, {Heinzelmann}, {Henri}, {Hermann}, {Hinton}, {Hofmann}, {Hoischen},
  {Holch}, {Holler}, {Horns}, {Huber}, {Iwasaki}, {Jamrozy}, {Jankowsky},
  {Jankowsky}, {Jardin-Blicq}, {Jung-Richardt}, {Kastendieck},
  {Katarzy{\'n}ski}, {Katsuragawa}, {Katz}, {Khangulyan}, {Kh{\'e}lifi},
  {King}, {Klepser}, {Klu{\'z}niak}, {Komin}, {Kosack}, {Kostunin}, {Kraus},
  {Lamanna}, {Lau}, {Lemi{\`e}re}, {Lemoine-Goumard}, {Lenain}, {Leser},
  {Levy}, {Lohse}, {Lypova}, {Mackey}, {Majumdar}, {Malyshev}, {Marandon},
  {Marcowith}, {Mares}, {Mariaud}, {Mart{\'\i}-Devesa}, {Marx}, {Maurin},
  {Meintjes}, {Mitchell}, {Moderski}, {Mohamed}, {Mohrmann}, {Moore}, {Moulin},
  {Muller}, {Murach}, {Nakashima}, {de Naurois}, {Ndiyavala}, {Niederwanger},
  {Niemiec}, {Oakes}, {O'Brien}, {Odaka}, {Ohm}, {de Ona Wilhelmi},
  {Ostrowski}, {Oya}, {Panter}, {Parsons}, {Perennes}, {Petrucci}, {Peyaud},
  {Piel}, {Pita}, {Poireau}, {Priyana Noel}, {Prokhorov}, {Prokoph},
  {P{\"u}hlhofer}, {Punch}, {Quirrenbach}, {Raab}, {Rauth}, {Reimer}, {Reimer},
  {Remy}, {Renaud}, {Rieger}, {Rinchiuso}, {Romoli}, {Rowell}, {Rudak},
  {Ruiz-Velasco}, {Sahakian}, {Saito}, {Sanchez}, {Santangelo}, {Sasaki},
  {Schlickeiser}, {Sch{\"u}ssler}, {Schulz}, {Schutte}, {Schwanke},
  {Schwemmer}, {Seglar-Arroyo}, {Senniappan}, {Seyffert}, {Shafi},
  {Shiningayamwe}, {Simoni}, {Sinha}, {Sol}, {Specovius}, {Spir-Jacob},
  {Stawarz}, {Steenkamp}, {Stegmann}, {Steppa}, {Takahashi}, {Tavernier},
  {Taylor}, {Terrier}, {Tiziani}, {Tluczykont}, {Trichard}, {Tsirou}, {Tsuji},
  {Tuffs}, {Uchiyama}, {van der Walt}, {van Eldik}, {van Rensburg}, {van
  Soelen}, {Vasileiadis}, {Veh}, {Venter}, {Vincent}, {Vink}, {Voisin},
  {V{\"o}lk}, {Vuillaume}, {Wadiasingh}, {Wagner}, {White}, {Wierzcholska},
  {Yang}, {Yoneda}, {Zacharias}, {Zanin}, {Zdziarski}, {Zech}, {Ziegler},
  {Zorn}, \& {{\.Z}ywucka}}]{2020Natur.582..356H}
{H.~E.~S.~S. Collaboration}, {Abdalla}, H., {Adam}, R., {et~al.} 2020, \nat,
  582, 356, \dodoi{10.1038/s41586-020-2354-1}

\bibitem[{{Harris} \& {Krawczynski}(2006)}]{2006ARA&A..44..463H}
{Harris}, D.~E., \& {Krawczynski}, H. 2006, \araa, 44, 463,
  \dodoi{10.1146/annurev.astro.44.051905.092446}

\bibitem[{{HI4PI Collaboration} {et~al.}(2016){HI4PI Collaboration}, {Ben
  Bekhti}, {Fl{\"o}er}, {Keller}, {Kerp}, {Lenz}, {Winkel}, {Bailin},
  {Calabretta}, {Dedes}, {Ford}, {Gibson}, {Haud}, {Janowiecki}, {Kalberla},
  {Lockman}, {McClure-Griffiths}, {Murphy}, {Nakanishi}, {Pisano}, \&
  {Staveley-Smith}}]{2016A&A...594A.116H}
{HI4PI Collaboration}, {Ben Bekhti}, N., {Fl{\"o}er}, L., {et~al.} 2016, \aap,
  594, A116, \dodoi{10.1051/0004-6361/201629178}

\bibitem[{{Lott} {et~al.}(2012){Lott}, {Escande}, {Larsson}, \&
  {Ballet}}]{2012A&A...544A...6L}
{Lott}, B., {Escande}, L., {Larsson}, S., \& {Ballet}, J. 2012, \aap, 544, A6,
  \dodoi{10.1051/0004-6361/201218873}

\bibitem[{{Mack} {et~al.}(1997){Mack}, {Klein}, {O'Dea}, \&
  {Willis}}]{1997A&AS..123..423M}
{Mack}, K.~H., {Klein}, U., {O'Dea}, C.~P., \& {Willis}, A.~G. 1997, \aaps,
  123, 423, \dodoi{10.1051/aas:1997166}

\bibitem[{{Massaro} {et~al.}(2004){Massaro}, {Perri}, {Giommi}, \&
  {Nesci}}]{2004A&A...413..489M}
{Massaro}, E., {Perri}, M., {Giommi}, P., \& {Nesci}, R. 2004, \aap, 413, 489,
  \dodoi{10.1051/0004-6361:20031558}

\bibitem[{{Migliori} {et~al.}(2011){Migliori}, {Grandi}, {Torresi}, {Dermer},
  {Finke}, {Celotti}, {Mukherjee}, {Errando}, {Gargano}, {Giordano}, \&
  {Giroletti}}]{2011A&A...533A..72M}
{Migliori}, G., {Grandi}, P., {Torresi}, E., {et~al.} 2011, \aap, 533, A72,
  \dodoi{10.1051/0004-6361/201116808}

\bibitem[{{Mukherjee} {et~al.}(2002){Mukherjee}, {Halpern}, {Mirabal}, \&
  {Gotthelf}}]{2002ApJ...574..693M}
{Mukherjee}, R., {Halpern}, J., {Mirabal}, N., \& {Gotthelf}, E.~V. 2002, \apj,
  574, 693, \dodoi{10.1086/340999}

\bibitem[{{NASA/IPAC Extragalactic Database (NED)}(2019)}]{NED}
{NASA/IPAC Extragalactic Database (NED)}. 2019, NASA/IPAC Extragalactic
  Database (NED),  IPAC, \dodoi{10.26132/NED1}

\bibitem[{{Nolan} {et~al.}(2012){Nolan}, {Abdo}, {Ackermann}, {Ajello},
  {Allafort}, {Antolini}, {Atwood}, {Axelsson}, {Baldini}, {Ballet},
  {Barbiellini}, {Bastieri}, {Bechtol}, {Belfiore}, {Bellazzini}, {Berenji},
  {Bignami}, {Blandford}, {Bloom}, {Bonamente}, {Bonnell}, {Borgland},
  {Bottacini}, {Bouvier}, {Brandt}, {Bregeon}, {Brigida}, {Bruel}, {Buehler},
  {Burnett}, {Buson}, {Caliandro}, {Cameron}, {Campana}, {Ca{\~n}adas},
  {Cannon}, {Caraveo}, {Casandjian}, {Cavazzuti}, {Ceccanti}, {Cecchi},
  {{\c{C}}elik}, {Charles}, {Chekhtman}, {Cheung}, {Chiang}, {Chipaux},
  {Ciprini}, {Claus}, {Cohen-Tanugi}, {Cominsky}, {Conrad}, {Corbet}, {Cutini},
  {D'Ammando}, {Davis}, {de Angelis}, {DeCesar}, {DeKlotz}, {De Luca}, {den
  Hartog}, {de Palma}, {Dermer}, {Digel}, {Silva}, {Drell}, {Drlica-Wagner},
  {Dubois}, {Dumora}, {Enoto}, {Escande}, {Fabiani}, {Falletti}, {Favuzzi},
  {Fegan}, {Ferrara}, {Focke}, {Fortin}, {Frailis}, {Fukazawa}, {Funk},
  {Fusco}, {Gargano}, {Gasparrini}, {Gehrels}, {Germani}, {Giebels},
  {Giglietto}, {Giommi}, {Giordano}, {Giroletti}, {Glanzman}, {Godfrey},
  {Grenier}, {Grondin}, {Grove}, {Guillemot}, {Guiriec}, {Gustafsson},
  {Hadasch}, {Hanabata}, {Harding}, {Hayashida}, {Hays}, {Hill}, {Horan},
  {Hou}, {Hughes}, {Iafrate}, {Itoh}, {J{\'o}hannesson}, {Johnson}, {Johnson},
  {Johnson}, {Johnson}, {Kamae}, {Katagiri}, {Kataoka}, {Katsuta}, {Kawai},
  {Kerr}, {Kn{\"o}dlseder}, {Kocevski}, {Kuss}, {Lande}, {Landriu},
  {Latronico}, {Lemoine-Goumard}, {Lionetto}, {Llena Garde}, {Longo},
  {Loparco}, {Lott}, {Lovellette}, {Lubrano}, {Madejski}, {Marelli}, {Massaro},
  {Mazziotta}, {McConville}, {McEnery}, {Mehault}, {Michelson}, {Minuti},
  {Mitthumsiri}, {Mizuno}, {Moiseev}, {Mongelli}, {Monte}, {Monzani},
  {Morselli}, {Moskalenko}, {Murgia}, {Nakamori}, {Naumann-Godo}, {Norris},
  {Nuss}, {Nymark}, {Ohno}, {Ohsugi}, {Okumura}, {Omodei}, {Orlando}, {Ormes},
  {Ozaki}, {Paneque}, {Panetta}, {Parent}, {Perkins}, {Pesce-Rollins},
  {Pierbattista}, {Pinchera}, {Piron}, {Pivato}, {Porter}, {Racusin},
  {Rain{\`o}}, {Rando}, {Razzano}, {Razzaque}, {Reimer}, {Reimer}, {Reposeur},
  {Ritz}, {Rochester}, {Romani}, {Roth}, {Rousseau}, {Ryde}, {Sadrozinski},
  {Salvetti}, {Sanchez}, {Saz Parkinson}, {Sbarra}, {Scargle}, {Schalk},
  {Sgr{\`o}}, {Shaw}, {Shrader}, {Siskind}, {Smith}, {Spandre}, {Spinelli},
  {Stephens}, {Strickman}, {Suson}, {Tajima}, {Takahashi}, {Takahashi},
  {Tanaka}, {Thayer}, {Thayer}, {Thompson}, {Tibaldo}, {Tibolla}, {Tinebra},
  {Tinivella}, {Torres}, {Tosti}, {Troja}, {Uchiyama}, {Vandenbroucke}, {Van
  Etten}, {Van Klaveren}, {Vasileiou}, {Vianello}, {Vitale}, {Waite},
  {Wallace}, {Wang}, {Werner}, {Winer}, {Wood}, {Wood}, {Wood}, {Yang}, \&
  {Zimmer}}]{2012ApJS..199...31N}
{Nolan}, P.~L., {Abdo}, A.~A., {Ackermann}, M., {et~al.} 2012, \apjs, 199, 31,
  \dodoi{10.1088/0067-0049/199/2/31}

\bibitem[{{Perley} {et~al.}(1984){Perley}, {Bridle}, \&
  {Willis}}]{1984ApJS...54..291P}
{Perley}, R.~A., {Bridle}, A.~H., \& {Willis}, A.~G. 1984, \apjs, 54, 291,
  \dodoi{10.1086/190931}

\bibitem[{{Persic} \& {Rephaeli}(2019)}]{2019MNRAS.490.1489P}
{Persic}, M., \& {Rephaeli}, Y. 2019, \mnras, 490, 1489,
  \dodoi{10.1093/mnras/stz2527}

\bibitem[{{Sambruna} {et~al.}(2004){Sambruna}, {Gliozzi}, {Donato},
  {Tavecchio}, {Cheung}, \& {Mushotzky}}]{2004A&A...414..885S}
{Sambruna}, R.~M., {Gliozzi}, M., {Donato}, D., {et~al.} 2004, \aap, 414, 885,
  \dodoi{10.1051/0004-6361:20031657}

\bibitem[{{Saunders} {et~al.}(1981){Saunders}, {Baldwin}, {Pooley}, \&
  {Warner}}]{1981MNRAS.197..287S}
{Saunders}, R., {Baldwin}, J.~E., {Pooley}, G.~G., \& {Warner}, P.~J. 1981,
  \mnras, 197, 287, \dodoi{10.1093/mnras/197.2.287}

\bibitem[{{Takeuchi} {et~al.}(2012){Takeuchi}, {Kataoka}, {Stawarz},
  {Takahashi}, {Maeda}, {Nakamori}, {Cheung}, {Celotti}, {Tanaka}, \&
  {Takahashi}}]{2012ApJ...749...66T}
{Takeuchi}, Y., {Kataoka}, J., {Stawarz}, {\L}., {et~al.} 2012, \apj, 749, 66,
  \dodoi{10.1088/0004-637X/749/1/66}

\bibitem[{{Urry} \& {Padovani}(1995)}]{1995PASP..107..803U}
{Urry}, C.~M., \& {Padovani}, P. 1995, \pasp, 107, 803, \dodoi{10.1086/133630}

\bibitem[{{Waggett} {et~al.}(1977){Waggett}, {Warner}, \&
  {Baldwin}}]{1977MNRAS.181..465W}
{Waggett}, P.~C., {Warner}, P.~J., \& {Baldwin}, J.~E. 1977, \mnras, 181, 465,
  \dodoi{10.1093/mnras/181.3.465}

\bibitem[{{Wegner} {et~al.}(2003){Wegner}, {Bernardi}, {Willmer}, {da Costa},
  {Alonso}, {Pellegrini}, {Maia}, {Chaves}, \&
  {Rit{\'e}}}]{2003AJ....126.2268W}
{Wegner}, G., {Bernardi}, M., {Willmer}, C.~N.~A., {et~al.} 2003, \aj, 126,
  2268, \dodoi{10.1086/378959}

\bibitem[{{Xue} {et~al.}(2017){Xue}, {Zhang}, {Cui}, {Liang}, \&
  {Zhang}}]{2017RAA....17...90X}
{Xue}, Z.-W., {Zhang}, J., {Cui}, W., {Liang}, E.-W., \& {Zhang}, S.-N. 2017,
  Research in Astronomy and Astrophysics, 17, 090,
  \dodoi{10.1088/1674-4527/17/9/90}

\bibitem[{{Zhang} {et~al.}(2010){Zhang}, {Bai}, {Chen}, \&
  {Liang}}]{2010ApJ...710.1017Z}
{Zhang}, J., {Bai}, J.~M., {Chen}, L., \& {Liang}, E. 2010, \apj, 710, 1017,
  \dodoi{10.1088/0004-637X/710/2/1017}

\bibitem[{{Zhang} {et~al.}(2009){Zhang}, {Bai}, {Chen}, \&
  {Yang}}]{2009ApJ...701..423Z}
{Zhang}, J., {Bai}, J.~M., {Chen}, L., \& {Yang}, X. 2009, \apj, 701, 423,
  \dodoi{10.1088/0004-637X/701/1/423}

\bibitem[{{Zhang} {et~al.}(2018){Zhang}, {Du}, {Guo}, {Zhang}, {Chen}, {Liang},
  \& {Zhang}}]{2018ApJ...858...27Z}
{Zhang}, J., {Du}, S.-s., {Guo}, S.-C., {et~al.} 2018, \apj, 858, 27,
  \dodoi{10.3847/1538-4357/aab9b2}

\bibitem[{{Zhang} {et~al.}(2012){Zhang}, {Liang}, {Zhang}, \&
  {Bai}}]{2012ApJ...752..157Z}
{Zhang}, J., {Liang}, E.-W., {Zhang}, S.-N., \& {Bai}, J.~M. 2012, \apj, 752,
  157, \dodoi{10.1088/0004-637X/752/2/157}

\end{thebibliography}
\bibliographystyle{aasjournal}

\clearpage

\begin{figure}
 \centering
   \includegraphics[angle=0,scale=0.60]{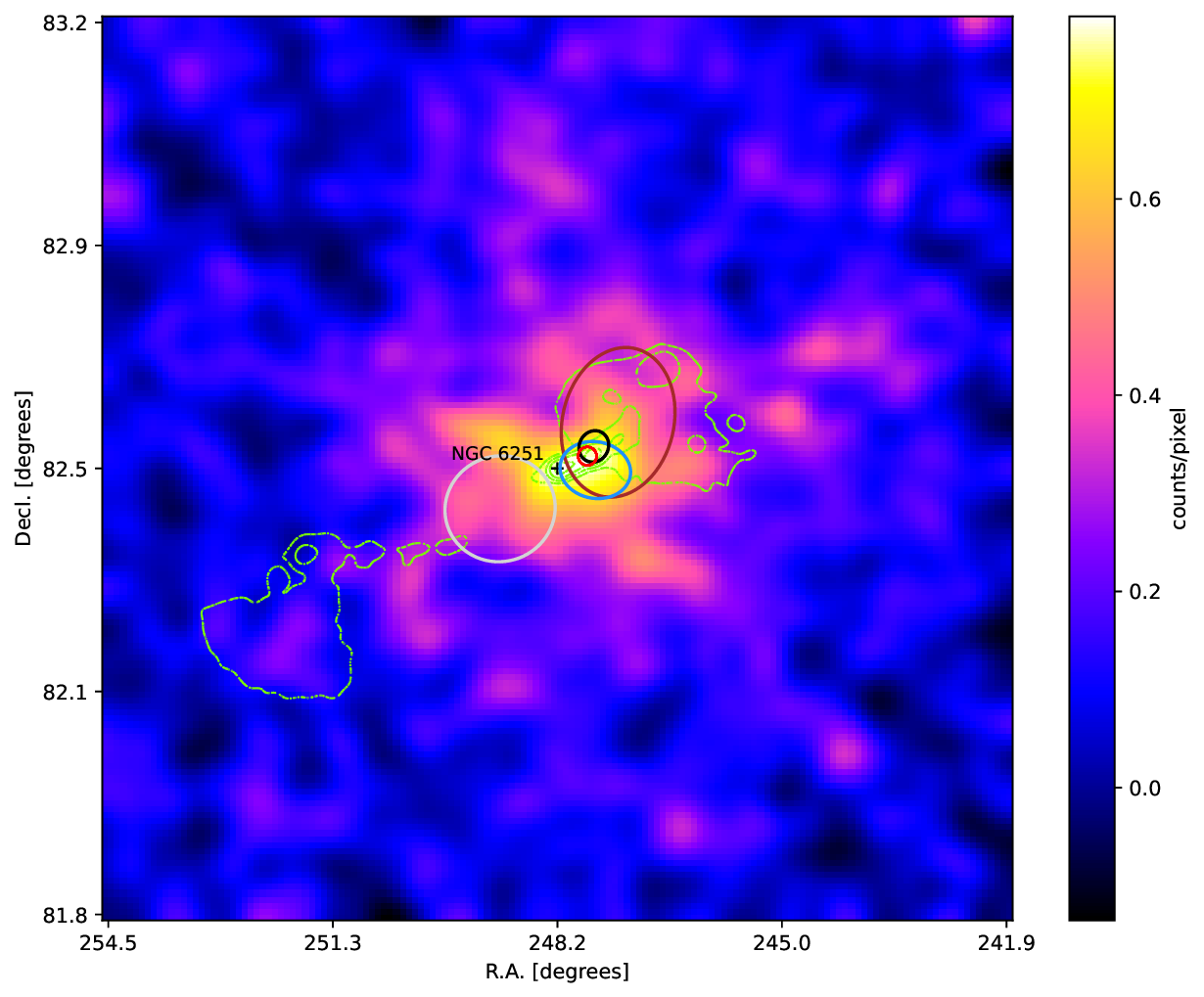}
\caption{$1\degr.5\times1\degr.5$ $\gamma$-ray counts map of 4FGL J1630.6+8234 in the 0.1--300 GeV band. The color bar represents the counts in each pixel with a pixel size of $0^\circ.01$. The map has been smoothed with a Gaussian kernel of $0^\circ.25$. The green contours denote the large-scale radio structure of NGC 6251 obtained with the WSRT observations \citep{1997A&AS..123..423M}. The black cross represents the position of the radio core in NGC 6251. The red circle indicates the 95\% error circle of the re-estimated best-fit position for 4FGL J1630.6+8234 obtained with the $\sim$15-yr Fermi-LAT observation data in this paper. The 95\% error ellipses of 4FGL J1630.6+8234 from 1FGL to 4FGL (marked respectively with gray, brown, blue, and black ellipses) are also presented for comparison.}\label{countsmap}
\end{figure}

\clearpage

\begin{table}[ht!]
\centering
\caption{Test Result for Different Spatial Templates}
\begin{threeparttable} 
\begin{tabular}{lcccccc}
        \hline\hline
        Model & $\Gamma_{\gamma}$ & $\beta$ & Energy Flux & TS & log(Likelihood)     \\
         &  &  & [10$^{-12}$ erg cm$^{-2}$ s$^{-1}$] &  &  \\ \hline
PS\tnote{a} & $2.36\pm0.03$ & $0.04\pm0.02$ & $13.40\pm0.51$ & 2619.34 &  219314.98   \\ \hline
RMT &  &  &  & 2697.62 & 219354.12  \\
Core\tnote{a} & $1.80\pm0.13$ & $0.28\pm0.06$ & $4.86\pm0.61$ & 508.55  &   \\
NW lobe\tnote{b} & $2.54\pm0.06$ & & $7.67\pm0.76$ & 600.31  &  \\
SE lobe\tnote{b} & $7.01\pm0.19$ &  & $0.96\pm0.23$ & 30.96  &  \\
\hline\hline
\end{tabular}  
\begin{tablenotes}   
        \footnotesize               
         \item[a] The log-parabola spectral form.  
         \item[b] The power-law spectral form.
\end{tablenotes}
\end{threeparttable} 
\label{tab:model}
\end{table}

\begin{table}[ht!]
\centering
\caption{Parameters for SED Fitting}
\begin{threeparttable} 
\begin{tabular}{lcc}
        \hline\hline
        Parameters & Core & NW lobe \\
         & syn+SSC & syn+SSC+IC/CMB  \\ \hline
$R$(cm) & $8.6\times 10^{16}$ & $5.7\times 10^{23}$ \\
$\delta=\Gamma$ & 3.4 & 1  \\
$B$(G) & 0.04 & $3.2\times 10^{-7}$   \\
$\gamma_{\min}$ & 200 & 100 \\
$\gamma_{\rm b}$ & $6.4\times 10^{3}$ & $3\times 10^{5}$\\
$\gamma_{\max}$ & $1.9\times 10^{5}$ & $3.5\times 10^{6}$ \\
$N_0$(cm$^{-3}$) & $3.8\times 10^{5}$ & $7\times 10^{-7}$\\
$p_1$ & 2.64 & 2.4\\
$p_2$ & 3.52 & 4\\
\hline\hline
\end{tabular}   
\end{threeparttable} 
\label{tab:SED}
\end{table}

\clearpage
\appendix
\section{Fermi-LAT Observations} 
\subsection{\texorpdfstring{$\gamma$}{gamma}-ray Counts map} 

\begin{figure}
 \centering
   \includegraphics[angle=0,scale=0.40]{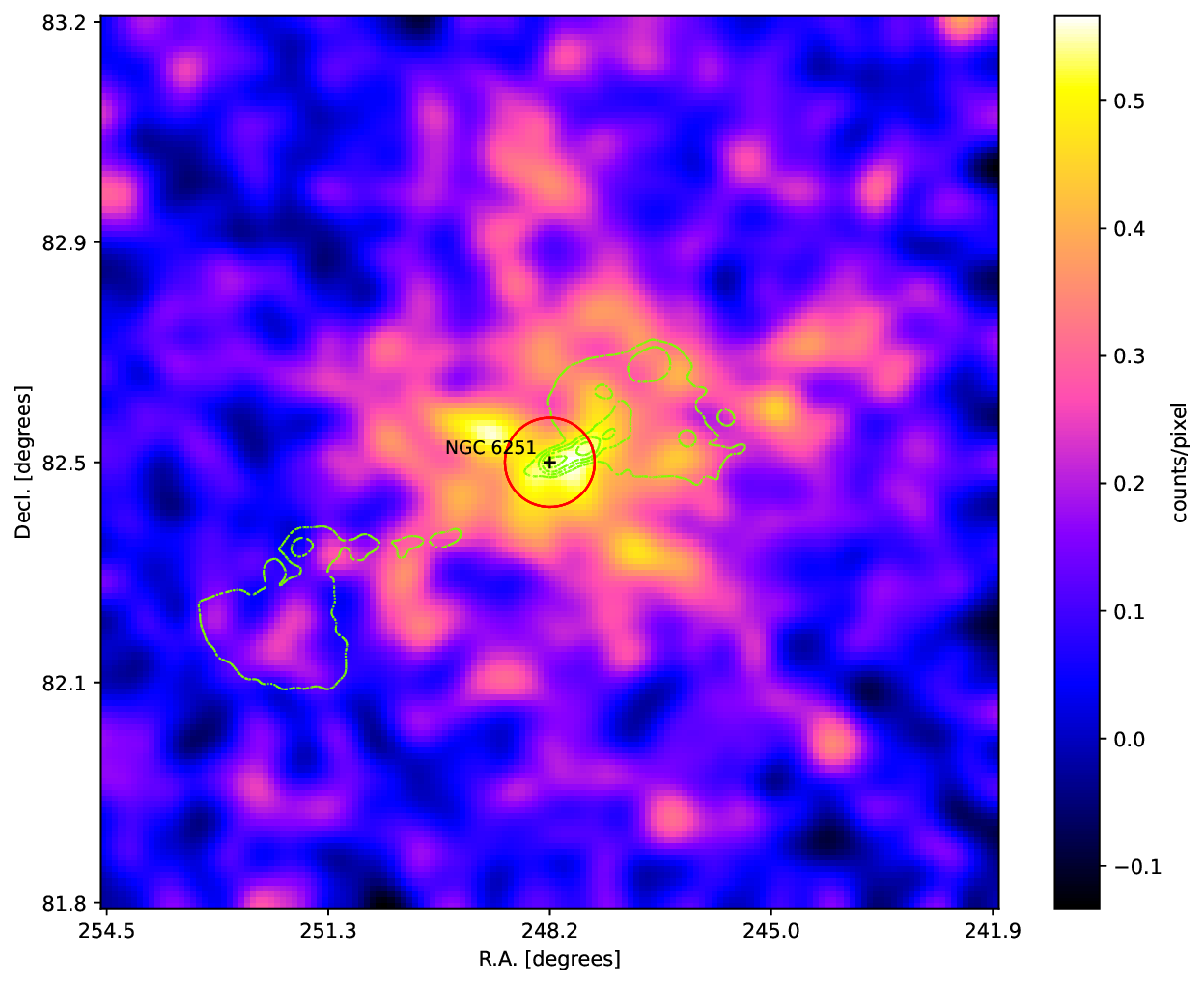}
     \includegraphics[angle=0,scale=0.40]{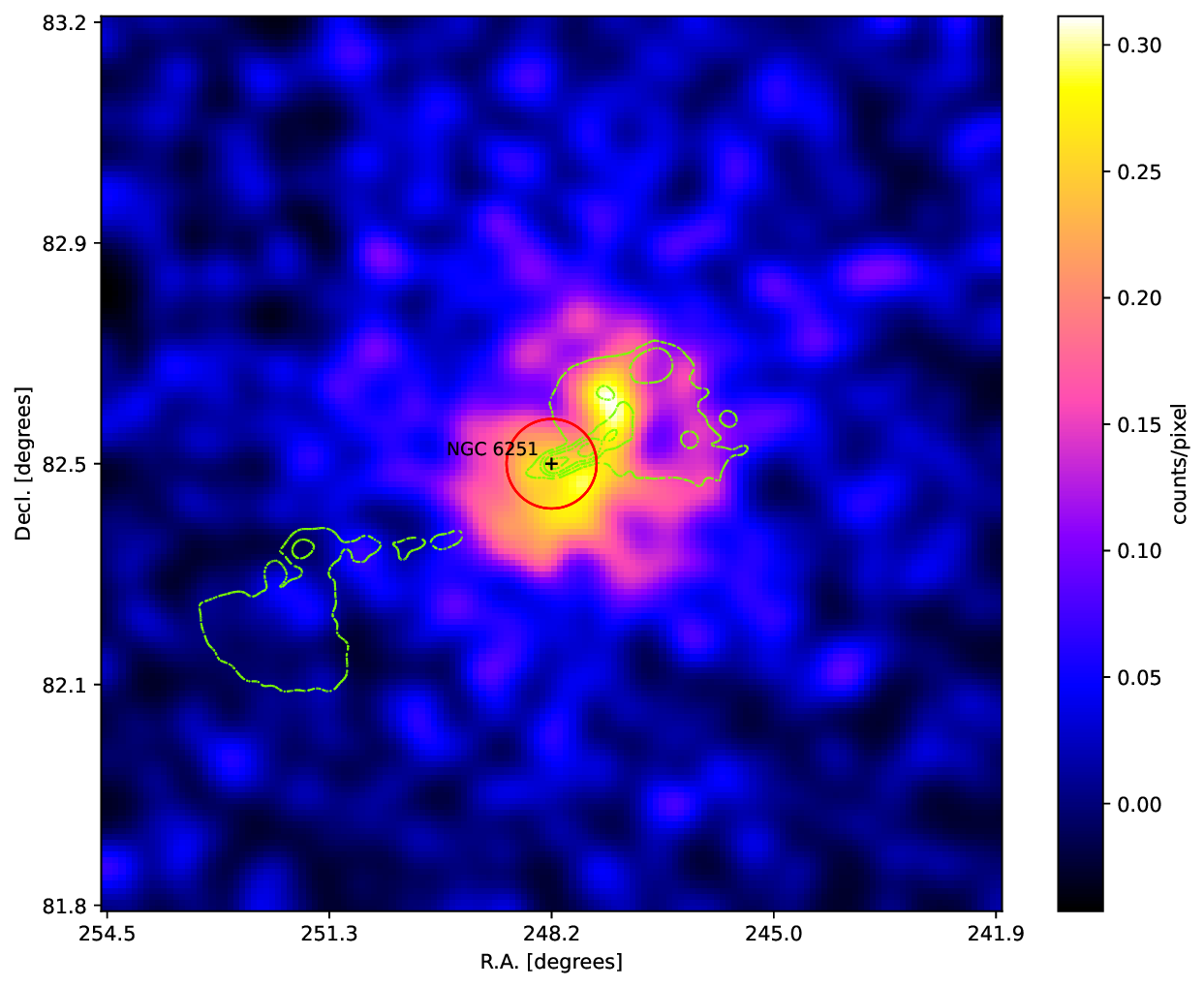}
       \includegraphics[angle=0,scale=0.40]{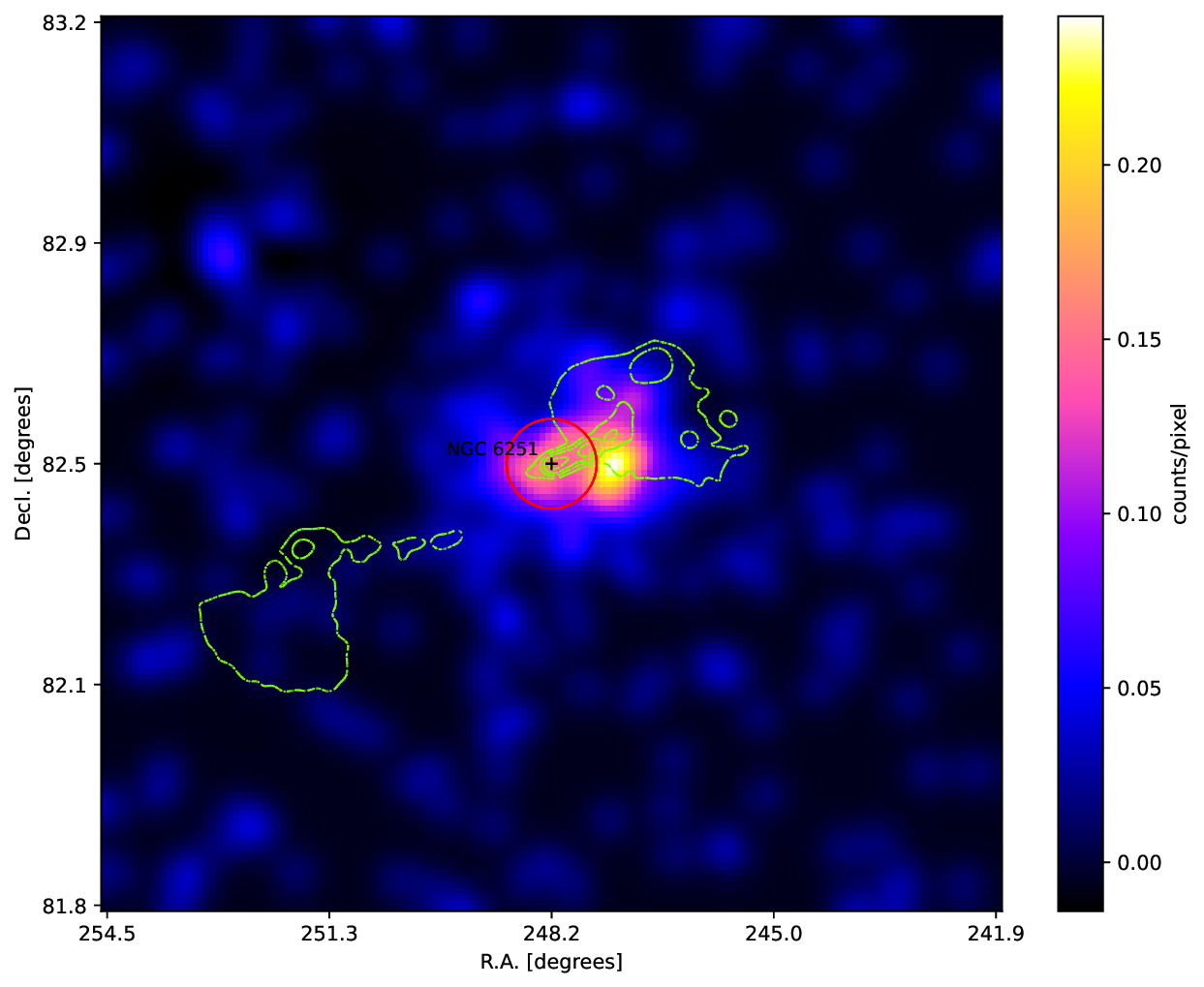}
         \includegraphics[angle=0,scale=0.40]{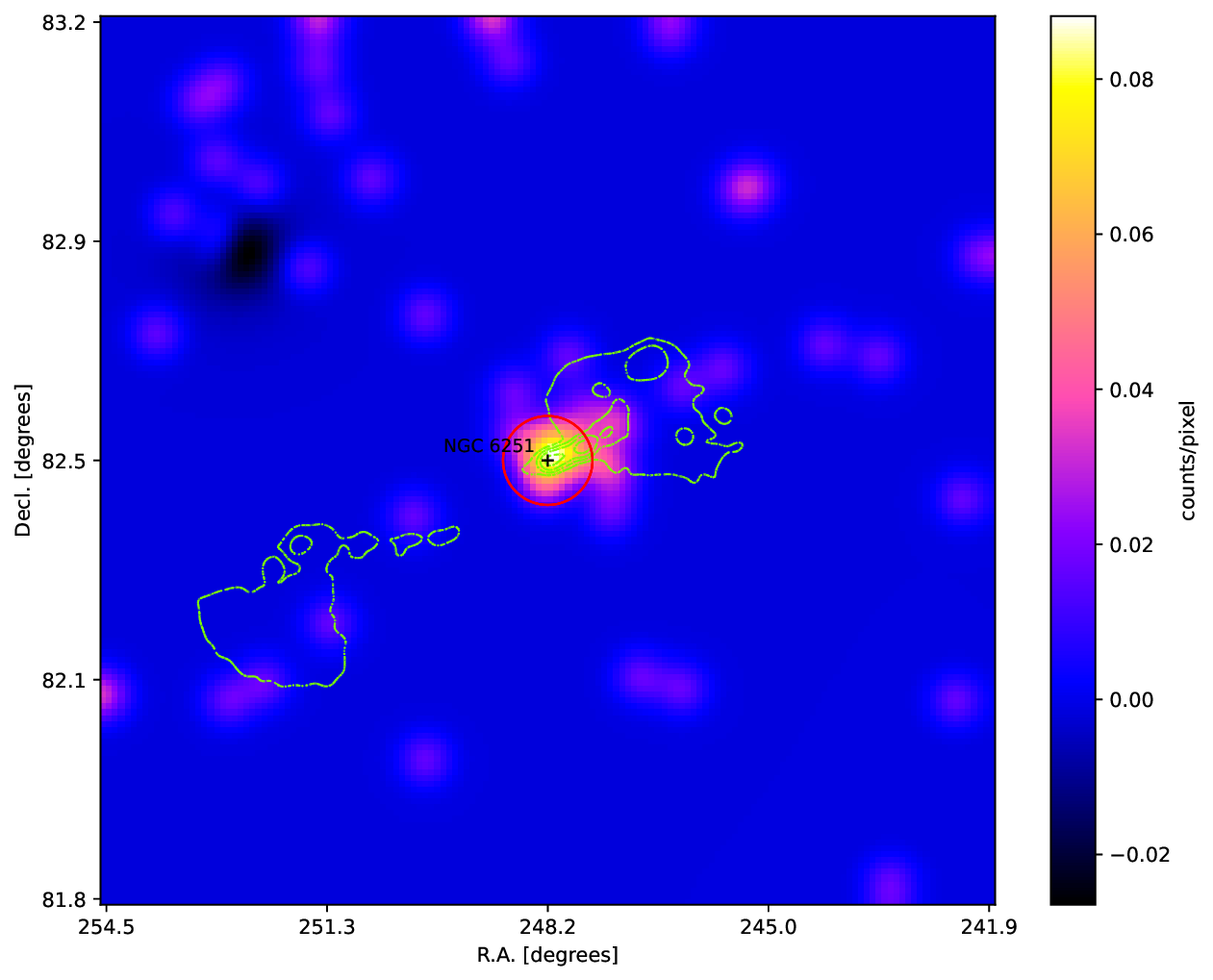}
\caption{The $1\degr.5\times1\degr.5$ $\gamma$-ray counts map of 4FGL J1630.6+8234 in different energy bands. The top left, top right, bottom left, and bottom right panels are in the energy bands of 0.1--3 GeV, 1--3 GeV, 3--10 GeV, and 10--300 GeV, respectively. The symbols remain unchanged from those in Figure \ref{countsmap}, with the exception of the red circle. Here the red circle represents a region centered on the position of the radio core with a radius of $4^{\prime}.5$. Note that the red circle is subtracted and not included for the NW lobe template in the RMT model.}\label{countsmap4}
\end{figure}

To generate the $\gamma$-ray counts map of 4FGL J1630.6+8234 in the 0.1--300 GeV band, we initially exclude the source 4FGL J1630.6+8234 from the XML file and retain only the background sources to create the background counts map. We then subtract the background counts from the total counts to obtain the $\gamma$-ray counts map of the target source, as depicted in Figure \ref{countsmap}. The size of each pixel in $\gamma$-ray counts maps is $0^\circ.01$. The value of color bar represents the number of $\gamma$-ray photon counts in each pixel. To further generate the $\gamma$-ray counts maps in different energy bands, we first perform data fitting by releasing only the normalization parameter of the target while keeping other model parameters fixed at their best-fit values obtained during the production of the $\gamma$-ray counts map for the 0.1--300 GeV band. Subsequently, we follow the same procedure as described above to obtain the $\gamma$-ray counts maps in different energy bands, as shown in Figure \ref{countsmap4}. 

\subsection{To Estimate the \texorpdfstring{$\gamma$}{gamma}-ray Variability} 

To quantify the variability of $\gamma$-ray light curves, we calculate $\chi^2$ and the associated probability $p(\chi^2)=1-p(>\chi^2)$ (\citealp{2022MNRAS.515.1723C} and references therein) of the long-term $\gamma$-ray light curves,
\begin{equation}
\label{eq1}
\langle{F}\rangle=\left[\sum\limits_{i=1}^N\frac{F_i}{\sigma^{2}_i}\right]\left[\sum\limits_{i=1}^N\frac{1}{\sigma^{2}_i}\right]^{-1}, 
\end{equation}
\begin{equation}
\label{eq2}
\chi^2=\sum\limits_{i=1}^N\frac{(F_i-\langle{F}\rangle)^2}{\sigma_{i}^2},
\end{equation}
where $N$ is the number of the data points, $\langle{F}\rangle$ is the weighted mean flux, ${F_i}$ and $\sigma_{i}$ are the flux and its error for the $i$th data point. However, none of the $\gamma$-ray light curves presented in this paper exhibits the significant variability at a confidence level exceeding 2$\sigma$.

\section{Swift-XRT Observations and Data Analysis}

The X-ray Telescope (XRT) on board the Neil Gehrels Swift Observatory (Swift, \citealp{2005SSRv..120..165B,2004AAS...20511601G}) has performed a total of 18 observations for NGC 6251 in the Photon Counting readout mode, spanning from April 2007 to October 2023. One observation was unable to capture the radio position of NGC 6251 due to it being outside the view field of Swift-XRT. For two other observations, poor quality data were obtained. Therefore, our analysis is based on a total of 15 observation data, as listed in Table \ref{tab:xrt}. The data are processed using the XRTDAS software package (v.3.7.0), where the software package was developed by the ASI Space Science Data Center and released by the NASA High Energy Astrophysics Archive Research Center in the HEASoft package (v.6.30.1). The calibration files from XRT CALDB (version 20220803) are used within \emph{xrtpipeline} to calibrate and clean the events. The individual XRT event files are merged together using the \emph{xselect} package, and then the average spectrum is created from the combined event file. Events for the spectral analysis are extracted from a circle centered on the radio core of NGC 6251 with a radius of 20$^{\prime\prime}$. The background is taken from an annulus with an inner and outer radii of 40$^{\prime\prime}$ and 80$^{\prime\prime}$. The ancillary response files, which is applied to correct the point-spread function losses and CCD defects, is generated with the \emph{xrtmkarf} task using the cumulative exposure map. The spectrum is grouped to ensure at least 20 counts per bin and the $\chi^2$ minimization technique is adopted for spectral analysis. The spectrum is fitted by a single PL with two absorption components, one is absorption at $z=0$ with the neutral hydrogen column density fixed at Galactic value of $N_{\rm H}^{\rm gal}=5.46\times10^{20}$ cm$^{-2}$ \citep{2016A&A...594A.116H}, the other is the intrinsic absorption of host galaxy ($N_{\rm H}^{\rm int}$) with column density set free. NGC 6251 was observed as a general observer target for an exposure of more than 10 ks on 2007 April 07. We first analyze this observational data and constrain the value of $N_{\rm H}^{\rm int}$, and then $N_{\rm H}^{\rm int}=7.60^{+7.82}_{-6.69}\times10^{20}$ cm$^{-2}$ is fixed throughout the following analysis. 

\begin{figure}
 \centering
   \includegraphics[angle=0,scale=0.45]{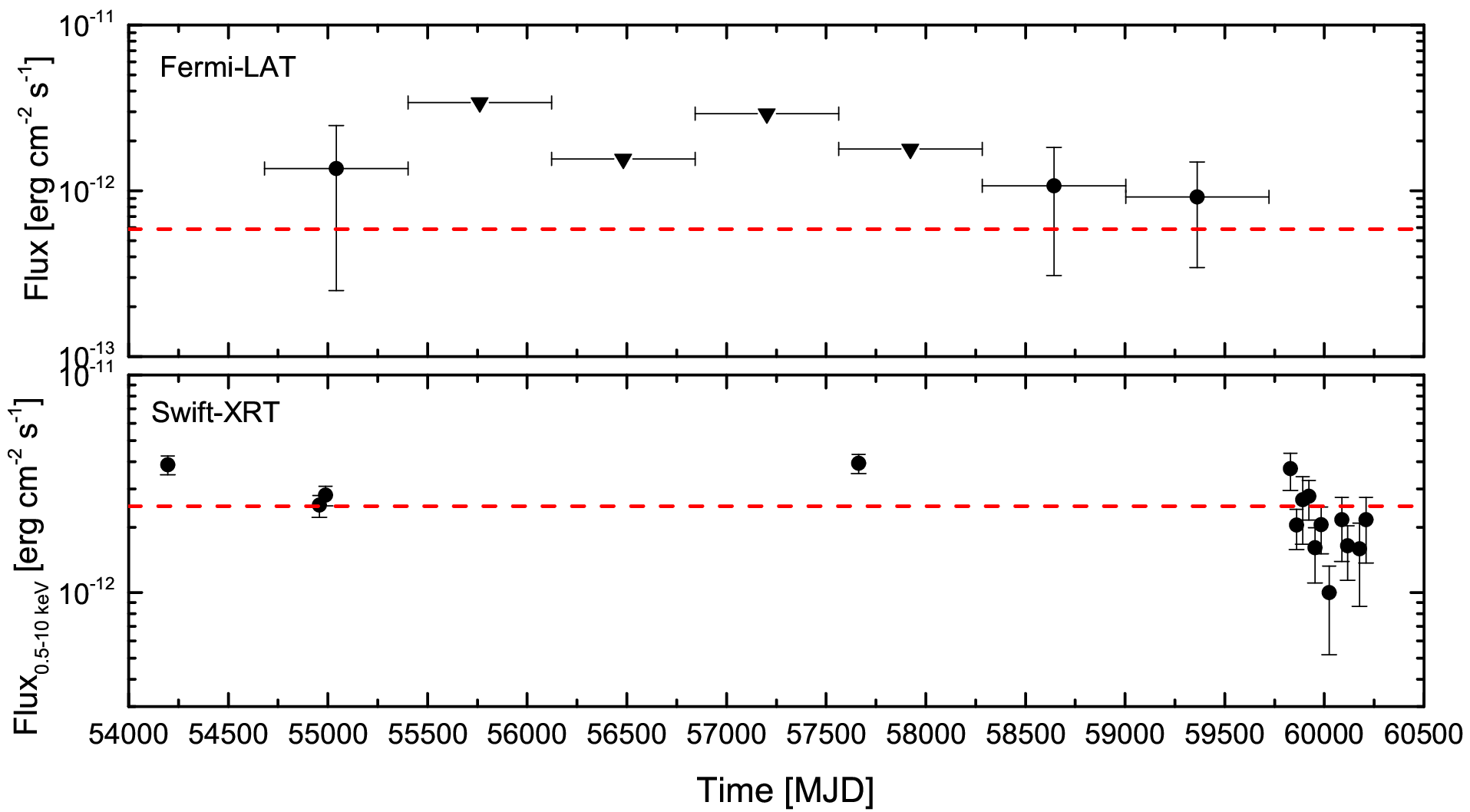}
\caption{Top Panel: the $\sim$15-yr $\gamma$-ray light curve in the 10--300 GeV band of 4FGL J1630.6+8234 when considering it as a PS, each time bin is 720 days and the triangles indicate TS$<$9 for this time bin. Bottom Panel: the X-ray light curve of NGC 6251, obtained by the Swift-XRT observation data. The horizontal red dashed line represents the average flux of all the data points, $\langle{F}\rangle_{\rm 0.5-10~keV}=2.50\times10^{-12}$ erg cm$^{-2}$ s$^{-1}$.}\label{two_lc}
\end{figure}

We generate the XRT spectra at other epochs following the procedures mentioned above. The derived photon index ($\Gamma_{\rm X}$) and corrected flux ($F_{\rm 0.5-10 keV}$) for the 15 observational epochs are given in Table \ref{tab:xrt}. The X-ray light curve obtained with the XRT observations is presented in Figure \ref{two_lc}. Using the above Equations (\ref{eq1}--\ref{eq2}), we obtain $\langle{F}\rangle_{\rm 0.5-10~keV}=2.50\times10^{-12}$ erg cm$^{-2}$ s$^{-1}$, $\chi^2=57.4$ (d.o.f$=N-1=14$) with $p=3.4\times10^{-7}$, indicating that there is variability of X-rays at a confidence level of $4.9\sigma$. 

We also produce a combined average spectrum by incorporating the latest 11 observational data spanning from 2022 September to 2023 October, which are all corrected values, as displayed in Figure \ref{SED}. It is worth noting that the first three Swift-XRT observational data were previously reported by \cite{2011A&A...533A..72M}, and our findings are consistent with them within the errors.

\clearpage

\begin{table}[ht!]
\centering
\caption{Analysis Results of the Swift-XRT Observation Data}
\begin{threeparttable} 
\begin{tabular}{lcccccc}
        \hline\hline
        obs. & exp. & norm & $\Gamma_{\rm X}$ & $\chi^2$(d.o.f) & $F_{\rm 0.5-10~keV}$    \\
        Date& [s] & [10$^{-4}$ ph keV$^{-1}$ cm$^{-2}$ s$^{-1}$] &  &  & [10$^{-12}$ erg cm$^{-2}$ s$^{-1}$] \\ \hline
2007--04--06&10958.413&$8.18^{+2.02}_{-1.57}$&$2.02^{+0.23}_{-0.21}$&16.6(24)&$3.88^{+0.37}_{-0.39}$\\
2009--05--05&5480.219&$5.57^{+0.81}_{-0.81}$&$2.07^{+0.22}_{-0.20}$&12.0(9)&$2.52^{+0.27}_{-0.31}$\\
2009--06--05&6438.284&$5.74^{+0.79}_{-0.78}$&$1.98^{+0.22}_{-0.20}$&12.7(12)&$2.80^{+0.27}_{-0.30}$\\
2016--10--01&6274.348&$6.77^{+0.97}_{-0.96}$&$1.78^{+0.21}_{-0.20}$&6.0(12)&$3.94^{+0.38}_{-0.41}$\\
2022--09--08&1824.569&$4.47^{+1.17}_{-1.14}$&$1.45^{+0.31}_{-0.30}$&15.3(16)&$3.72^{+0.64}_{-0.78}$\\
2022--10--08&1997.201&$4.94^{+1.30}_{-1.28}$&$2.20^{+0.49}_{-0.41}$&10.2(13)&$2.04^{+0.38}_{-0.46}$\\
2022--11--08&1119.854&$3.19^{+2.00}_{-1.57}$&$1.44^{+0.95}_{-0.72}$&6.5(9)&$2.68^{+0.73}_{-1.01}$\\
2022--12--08&1647.465&$6.05^{+1.46}_{-1.46}$&$2.06^{+0.36}_{-0.33}$&16.4(14)&$2.78^{+0.51}_{-0.62}$\\
2023--01--08&1746.515&$3.90^{+1.53}_{-1.43}$&$2.21^{+0.77}_{-0.58}$&8.3(10)&$1.61^{+0.38}_{-0.50}$\\
2023--02--08&1671.833&$4.62^{+1.39}_{-1.40}$&$2.10^{+0.70}_{-0.58}$&9.9(10)&$2.05^{+0.43}_{-0.54}$\\
2023--03--21&1029.789&$2.98^{+1.35}_{-1.37}$&$2.61^{+1.57}_{-0.83}$&11.0(13)&$1.00^{+0.32}_{-0.48}$\\
2023--05--24&904.465&$4.43^{+1.71}_{-1.69}$&$1.98^{+0.63}_{-0.53}$&6.6(12)&$2.17^{+0.57}_{-0.78}$\\
2023--06--21&1466.011&$4.34^{+1.35}_{-1.36}$&$2.36^{+0.65}_{-0.55}$&13.3(24)&$1.64^{+0.38}_{-0.50}$\\
2023--08--21&931.808&$2.63^{+1.28}_{-1.26}$&$1.75^{+0.69}_{-0.62}$&6.5(11)&$1.59^{+0.50}_{-0.72}$\\
2023--09--21&943.641&$3.98^{+1.70}_{-1.71}$&$1.85^{+0.72}_{-0.66}$&10.5(18)&$2.17^{+0.58}_{-0.79}$\\
\hline\hline
\end{tabular}
\end{threeparttable} 
\label{tab:xrt}
\end{table}

\end{document}